\documentclass[]{jkpaper}
\usepackage{tikz}
\usepackage{aas_macros}
\usetikzlibrary{decorations.markings, decorations.pathreplacing}
\usetikzlibrary{decorations.pathmorphing, patterns,shapes}
\usetikzlibrary{arrows.meta}

\newcommand{\OIST}{\raisebox{-0.08em}{\includegraphics[height=0.8em]{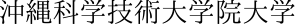}}}

\newtheorem*{lemma}{Lemma}

\title{Islands and Uhlmann phase:\texorpdfstring{\\}{ }Explicit recovery of classical information from evaporating black holes}
\author{Josh Kirklin}
\email{joshua.kirklin@oist.jp}
\institution{Qubits and Spacetime Unit,\texorpdfstring{\\}{ }Okinawa Institute of Science and Technology\texorpdfstring{\\}{ } \OIST}

\abstr{%
    Recent work has established a route towards the semiclassical validity of the Page curve, and so provided evidence that information escapes an evaporating black hole. However, a protocol to explicitly recover and make practical use of that information in the classical limit has not yet been given. In this paper, we describe such a protocol, showing that an observer may reconstruct the phase space of the black hole interior by measuring the Uhlmann phase of the Hawking radiation. The process of black hole formation and evaporation provides an invertible map between this phase space and the space of initial matter configurations. Thus, all classical information is explicitly recovered. We assume in this paper that replica wormholes contribute to the gravitational path integral.
}

\bibliography{refs}

\begin{document}

\maketitleandtoc

\section{Introduction}
\label{Section: Introduction}

Classical black holes are airtight, but semiclassical ones are leaky, gradually losing energy over time via the Hawking process~\cite{ParticleCreation,BlackHoleExplosions}. In the absence of any infalling matter, they will diminish in size until they reach the Planck scale, at which point higher order quantum gravitational effects take over. Before that point, Hawking's original calculations suggest that the entropy of the radiation emitted by the black hole will steadily grow~\cite{BreakdownOfPredictability}. But for the combined process of black hole formation and evaporation to be unitary, the entropy must instead follow the Page curve -- at a certain point (the Page time) it must stop growing and start decreasing, eventually becoming very small after the black hole is gone~\cite{PageCurve,PageCurve2}. The Page time takes place well before the black hole reaches the Planck scale, so there is tension between Hawking's picture and unitarity. If we believe quantum gravity is unitary\footnote{Quantum gravity does not have to be unitary, but most of the theoretical physics community seems to believe it is at the moment. One reason for this is the existence of AdS/CFT~\cite{AdSCFT1,AdSCFT2}, which shows that unitary theories of quantum gravity exist by construction.}, then this constitutes a paradox.

The resolution of this issue has remained beyond our reach for nearly half a century, but a flurry of recent progress indicates that we may be getting closer~\cite{Flurry0,Flurry1,Flurry2,Flurry3,Flurry4,Flurry5}. Apparently, we have been computing the entropy $S$ of the Hawking radiation in the wrong way. We should have been using the `island formula':
\begin{equation}
    S = \min\qty{\operatorname*{ext}_I\qty[\frac{\operatorname{Area}(\partial I)}{4G} + S_{\text{bulk}}(I\cup R)]}.
    \label{Equation: island formula}
\end{equation}
$R$ is a region of space containing all of the Hawking radiation, while $I$ is a separate region called the `island'. $\operatorname{Area}(\partial I)$ is the area of the boundary of the island,\footnote{This area term is appropriate for general relativity. In higher derivative theories of gravity, it is replaced by some other functional of the geometry of $\partial I$~\cite{HigherDerivative1,HigherDerivative2}.} $G$ is Newton's constant, and $S_{\text{bulk}}(I\cup R)$ is the so-called `bulk entropy' of the combined region $I\cup R$, i.e.\ the entropy of perturbative QFT fluctuations (including gravitons and any other field excitations) on top of a fixed background geometry. According to~\eqref{Equation: island formula}, we need to find the set of extrema of the functional $\operatorname{Area}(\partial I)/4G + S_{\text{bulk}}(I\cup R)$ with respect to choices of the island $I$. The entropy $S$ is then supposed to be given by the minimum value in this set.

Consider the island $I$ for which the minimum extremum is attained.\footnote{We will interchangably use $I$ to refer to either the free variable over which the extremisation is carried out in~\eqref{Equation: island formula}, or the island at the minimum extremum. In each instance, it should be clear from the context to which we are referring.} At first, this $I$ is empty, and $S$ reduces to the bulk entropy $S_{\text{bulk}}(R)$ of the Hawking radiation $R$. However, at the Page time a non-empty island appears in the black hole, and during the subsequent evaporation, $I$ continues to grow in size until it spans almost the entire black hole interior. The papers~\cite{Flurry0,Flurry1,Flurry2,Flurry3,Flurry4,Flurry5} argue that accounting for this in the island formula reproduces the Page curve, thus removing the tension between black hole evaporation and unitarity in quantum gravity. So, in principle, any information inside the black hole eventually escapes and becomes accessible to an external observer. 

Despite this progress, it has not been made clear exactly what form the escaped information takes in the classical limit, or where a classical observer should look for it. The problem is that the information has been heavily encrypted by the black hole formation and evaporation process, and no protocol for decrypting it has yet been firmly established. In this paper, we will describe such a protocol. 

In particular, we will explain how an observer can use the Hawking radiation to reconstruct the classical phase space, including the symplectic form and therefore all the Poisson brackets, of the fields located in the island $I$. Thus, at the Page time, the observer starts to learn about the black hole interior. If the observer waits a sufficiently long time, the island will grow large enough that all classical degrees of freedom in the interior of the black hole are accounted for in the island phase space. Then the dynamical map $\mathscr{S}$ giving the evolution from the initial phase space of classical collapsing matter configurations to the island phase space is invertible. The structure of a phase space determines the full set of measurements a classical observer can make, so the invertibility of $\mathscr{S}$ means that a late-time classical observer and an early-time classical observer have access to the exact same set of measurements. In this way all the classical information is explicitly recovered.

The entropy formula~\eqref{Equation: island formula} is inspired by similar formulae in AdS/CFT and holography~\cite{HRT1,HRT2,HRT3,HRT4,HRT5}, but it is supposed to hold in a general theory of quantum gravity. Similarly, the protocol presented here takes inspiration from the recent identification of the holographic dual of the bulk symplectic form~\cite{BLS1,BLS2,Kirklin4}. Suppose we are given a closed curve of holographic states $\ket{\lambda}$ in AdS/CFT, and a boundary subregion of interest $A$. By tracing over degrees of freedom in the complement of $A$, one obtains a closed curve of reduced states $\rho_A$ in $A$. For any such curve, one can compute a number $\gamma$ called the Uhlmann phase~\cite{Uhlmann1992,Uhlmann1991,Uhlmann:sp,Braunstein,Jozsa:1994,Bures,Helstrom}. This number depends \emph{only} on the curve of reduced states $\rho_A$ -- we will explain later the precise way in which it is defined. The key result of~\cite{Kirklin4} was the formula
\begin{equation}
    \gamma = \int_C\Theta,
    \label{Equation: holographic uhlmann}
\end{equation}
where $C$ is the curve of bulk field configurations dual to the states $\ket{\lambda}$, and $\Omega=\dd{\Theta}$ is the symplectic form of the entanglement wedge of $A$. It is convenient to view $\Theta$ as a connection, and $\Omega$ as the curvature of this connection. Thus, the entanglement wedge symplectic form is holographically dual to the curvature of the boundary Uhlmann phase.

The analogue of the entanglement wedge for an evaporating black hole is the domain of dependence of the island $I$, so one might expect that by measuring the curvature of the Uhlmann phase of the Hawking radiation one can reproduce the symplectic form for the fields on the island. This would imply that the space of possible Hawking radiation states is a symplectic manifold, isomorphic to the island phase space. The main point of the present work is to confirm that this expectation is correct by bringing together the insights that led to~\eqref{Equation: island formula} and~\eqref{Equation: holographic uhlmann}. Both of the derivations for these formulae make use the same kind of mathematical tool -- namely the replica trick in a gravitational path integral. To get~\eqref{Equation: island formula}, one has to account for so-called `replica wormholes' in the path integral. Our results will follow from similarly accounting for these wormholes. There has been a lot of discussion on whether or not including them is the right thing to do~\cite{QuestionWormholes1,QuestionWormholes2,QuestionWormholes3}, but we will just assume that it is and make no further comments. 

To summarise, we will show that the curvature of the Hawking radiation Uhlmann phase is equivalent to the symplectic form of the island $I$, and that at late times semiclassical evolution provides an invertible map $\mathscr{S}$ between the island phase space and the initial phase space. Together these make up a complete decryption protocol for recovering classical information. The rest of the paper starts in Section~\ref{Section: Islands} with a description of the set-up, and a brief review of how the contributions of replica wormholes in the replica trick lead to the island formula~\eqref{Equation: island formula}. Then, in Section~\ref{Section: Uhlmann phase}, we define the Uhlmann phase, and explain how to compute it for Hawking radiation using an extension of the replica trick. In Section~\ref{Section: Information recovery}, we show how to \changed{use this result to }reconstruct the island phase space, \changed[and we also exhibit the dynamical map $\mathscr{S}$]{and how to use the reconstruction to recover classical information from the black hole interior}. We finish in Section~\ref{Section: Conclusion} with a short discussion of the broader consequences of our results.

\section{Evaporating black hole and islands}
\label{Section: Islands}

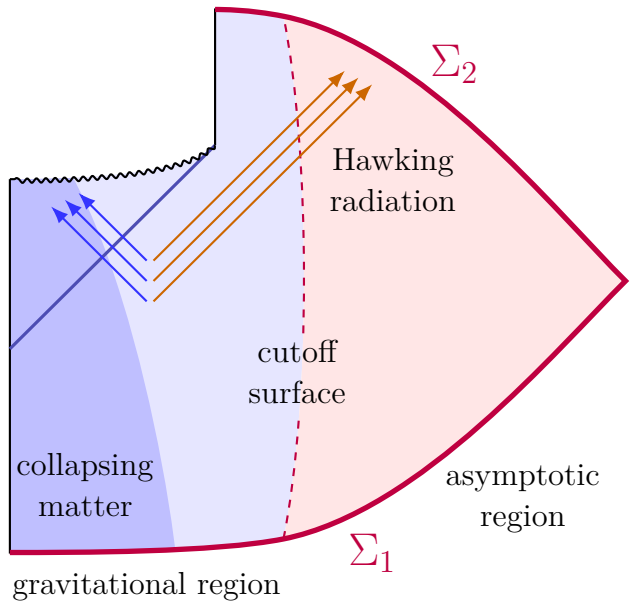
\begin{figure}
    \centering
    \begin{tikzpicture}[thick, scale=0.9]
        \fill[blue!10] (0,0) .. controls (1,0) and (3,0) .. (4,0.2)
            .. controls (4.4,2.2) and (4.4,5.9) .. (4,7.9)
            .. controls (3.5,8) and (3.1,8) .. (3,8)
            -- (3,6) 
            .. controls (2.5,5.5) and (0.5,5.5) .. (0,5.5)
            -- (0,0);
        \fill[red!10] (4,0.2) .. controls (6,0.6) and (8,3) .. (9,4)
            .. controls (8,5) and (6,7.5) .. (4,7.9)
            .. controls (4.4,5.9) and (4.4,2.2) .. (4,0.2);
        \begin{scope}
            \clip (0,0) .. controls (1,0) and (3,0) .. (4,0.2)
                .. controls (4.4,2.2) and (4.4,5.9) .. (4,7.9)
                .. controls (3.5,8) and (3.1,8) .. (3,8)
                -- (3,6) 
                .. controls (2.5,5.5) and (0.5,5.5) .. (0,5.5)
                -- (0,0);
            \fill[blue!25] (2.5,-1) .. controls (2.5,0.5) and (1.5,6) .. (0,6.5) -- (0,0);
        \end{scope}

        \draw (0,5.5) -- (0,0) .. controls (1,0) and (3,0) .. (4,0.2)
            .. controls (6,0.6) and (8,3) .. (9,4)
            .. controls (8,5) and (6,7.5) .. (4,7.9)
            .. controls (3.5,8) and (3.1,8) .. (3,8)
            -- (3,6);
        \draw[line width=2pt,purple] (0,0) .. controls (1,0) and (3,0) .. (4,0.2)
            .. controls (6,0.6) and (8,3) .. (9,4)
            .. controls (8,5) and (6,7.5) .. (4,7.9)
            .. controls (3.5,8) and (3.1,8) .. (3,8);
        \draw[very thick, blue!40!gray] (3,6) -- (0,3);
        \draw[decorate,decoration={snake,amplitude=0.8,segment length=3.5}] (3,6) .. controls (2.5,5.5) and (0.5,5.5) .. (0,5.5);
        \begin{scope}
            \clip (3,3.21) rectangle (5,8);
            \draw[dashed,purple] (4,0.2) .. controls (4.4,2.2) and (4.4,5.9) .. (4,7.9);
        \end{scope}
        \begin{scope}
            \clip (3,2.05) rectangle (5,0);
            \draw[dashed,purple] (4,0.2) .. controls (4.4,2.2) and (4.4,5.9) .. (4,7.9);
        \end{scope}

        \begin{scope}[-{Latex[]},orange!80!black]
            \draw (2.1,4) -- (5.1,7);
            \draw (2.1,3.7) -- (5.3,6.9);
            \draw (2.1,4.3) -- (4.9,7.1);
        \end{scope}
        \begin{scope}[-{Latex[]},blue!80!white]
            \draw (2,4) -- (0.8,5.2);
            \draw (2,3.7) -- (0.6,5.1);
            \draw (2,4.3) -- (1.0,5.3);
        \end{scope}

        \node[text width=3cm, align=center] at (1.1,1) {collapsing matter};
        \node at (2,-0.5) {gravitational region};
        \node[text width=3cm, align=center] at (7.5,0.8) {asymptotic region};
        \node[text width=2cm, align=center] at (4.2,2.66) {cutoff surface};
        \node[text width=2cm, align=center] at (5.6,5.5) {Hawking radiation};
        \node[purple] at (5.3,0) {\Large$\Sigma_1$};
        \node[purple] at (6.5,7.2) {\Large$\Sigma_2$};
    \end{tikzpicture}
    \caption{Spacetime is divided into asymptotic and gravitational regions, joined by a cutoff surface. Initially, the system is set up with a matter configuration that collapses to form a black hole. Then, the black hole evaporates, and all of the Hawking radiation is collected by the asymptotic region. (Note that we will continue to use this color scheme, in which the asymptotic region is red and the gravitational region is blue, throughout the paper.)}
    \label{Figure: set-up}
\end{figure}

The set-up we consider consists of a spacetime divided into a gravitational region and an asymptotic region, which are coupled together at a `cutoff surface'. We assume that gravitational effects only play a role in the gravitational region. This could be an exact assumption, so that the theories in the two regions are genuinely different, or it could just be an approximation. We form a black hole in the gravitational region, and then allow the asymptotic region to absorb the resulting Hawking radiation, until the black hole evaporates. This is illustrated in Figure~\ref{Figure: set-up}.\footnote{Our figures will depict spacetime as if it is asymptotically flat, but the precise behaviour of the fields in the asymptotic region is not particularly important for our purposes.} The spacetime depicted lies between two spacelike surfaces $\Sigma_1$ and $\Sigma_2$. We would like to think of them as Cauchy surfaces on which the initial and final states live, and indeed $\Sigma_1$ is a genuine Cauchy surface. But $\Sigma_2$ is not, because its domain of dependence does not include a large chunk of the spacetime, including the black hole region and the collapsing matter. As a consequence, the map representing the evolution from $\Sigma_1$ to $\Sigma_2$ is not invertible, so from a naive perspective classical information appears to have been lost down the black hole. This is the problem we intend to solve in this paper.

Let $\mathcal{H}_G,\mathcal{H}_R$ denote the Hilbert spaces of the gravitational region and the asymptotic region respectively, so that the total Hilbert space of the full system is $\mathcal{H}_G\otimes\mathcal{H}_R$.\footnote{This factorisation comes with two caveats. First, the algebra of observables of a subregion in a continuum QFT or gravitational theory is believed to generically be of Von Neumann type III~\cite{TypeIII}. Such algebras cannot be associated with tensor factors of the full Hilbert space. In this paper, we will assume that the theory under consideration has been regularised (perhaps on a lattice), so that the subsystems are no longer of type III. Second, even after such regularisation, the presence of gauge symmetries and edge modes means that the total physical Hilbert space is not a tensor product, but rather a direct sum of tensor products. Again, we will assume that the theory has been sufficiently regularised that we can isometrically embed the physical Hilbert space inside a `kinematic' Hilbert space of the desired form $\mathcal{H}_G\otimes\mathcal{H}_R$. We expect that our results are regularisation-independent, but it would be interesting and important to check this directly. Such a check is beyond the scope of the present paper, and would require an algebraic definition of Uhlmann phase.} During the evolution of the system, the Hawking radiation is collected in the part of the state in $\mathcal{H}_R$. For this reason, one can think of $\mathcal{H}_R$ as actually being the Hilbert space of the Hawking radiation itself.

\subsection{States and partition functions}

\begin{figure}
    \centering
    \begin{tikzpicture}[thick,scale=1.6]
        \fill[blue!15] (0,0) .. controls (1,-2) and (4,-2) .. (5,0)
            .. controls (4,2) and (1,2) .. (0,0);
        \begin{scope}
            \clip (0,0) .. controls (1,-2) and (4,-2) .. (5,0)
                .. controls (4,2) and (1,2) .. (0,0);
            \fill[red!15] (4,0) ellipse (1.5 and 2);
            \draw[purple,dashed] (4,0) ellipse (1.5 and 2);
        \end{scope}
        \draw (0,0) .. controls (1,-2) and (4,-2) .. (5,0)
            .. controls (4,2) and (1,2) .. (0,0);

        \node at (2.5,-1.8) {\Large$\mathcal{N}_-,\lambda_1$};
        \node at (2.5,1.8) {\Large$\mathcal{N}_+,\lambda_2^{*\mathrm{T}}$};

        \node[text width=3cm, align=center] at (3.7,0) {asymptotic region};
        \node[text width=3cm, align=center] at (1.3,0) {gravitational region};
    \end{tikzpicture}
    \caption{The manifold and fields for the partition function $Z(\lambda_2,\lambda_1)$ must obey the boundary conditions $\lambda_1$ in the past $\mathcal{N}_-$, and the complex conjugate and time reversal of $\lambda_2$ in the future $\mathcal{N}_+$.}
    \label{Figure: partition function boundary conditions}
\end{figure}
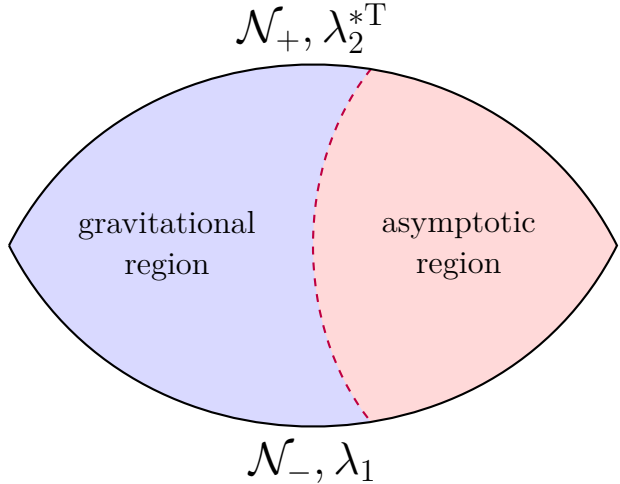

The states $\ket{\lambda}$ in $\mathcal{H}_G\otimes\mathcal{H}_R$ which we will consider essentially amount to the specification of a set of boundary conditions $\lambda$ for the fields on a surface. The inner product of such states is given by
\begin{equation}
    \braket{\lambda_2}{\lambda_1} = \frac{Z(\lambda_2,\lambda_1)}{\sqrt{Z(\lambda_2,\lambda_2)Z(\lambda_1,\lambda_1)}},
    \label{Equation: inner product}
\end{equation}
where $Z(\lambda_2,\lambda_1)$ is a Euclidean partition function. The denominator ensures that $\ket{\lambda}$ is normalised. In a semiclassical limit for the gravitational fields, we can approximate the partition function with
\begin{equation}
    -\log Z(\lambda_2,\lambda_1) = I_{\text{grav}}[\mathcal{M},g] - \log Z_{\text{QFT}}[\mathcal{M},g].
\end{equation}
Here, $\mathcal{M}$ is a manifold with metric $g$. $I_{\text{grav}}$ is the effective gravitational action of $\mathcal{M},g$, and $Z_{\text{QFT}}$ is a QFT partition function for fields on top of $\mathcal{M}$. The boundary $\mathcal{N}=\partial\mathcal{M}$ splits into two pieces: the `past' boundary $\mathcal{N}_-$, and the `future' boundary $\mathcal{N}_+$. All of the fields, including the metric $g$, have to obey the boundary conditions $\lambda_1$ on $\mathcal{N}_-$ and $\lambda_2^{*\mathrm{T}}$ on $\mathcal{N}_+$, where $\lambda_2^{*\mathrm{T}}$ denotes the complex conjugate and time reversal of $\lambda_2$. This is shown in Figure~\ref{Figure: partition function boundary conditions}. Because we are doing a saddlepoint calculation, the right hand side is supposed to be evaluated on the $\mathcal{M},g$ for which it is minimised.

Taking a semiclassical limit for the QFT fields as well, we can write
\begin{equation}
    -\log Z_{\text{QFT}}[\mathcal{M},g] = I_{\text{fields}}[\mathcal{M},\phi],
\end{equation}
where $I_{\text{fields}}$ is the effective action for the fields $\phi$ (which include the metric $g$). The full partition function is then given by
\begin{equation}
    -\log Z(\lambda_2,\lambda_1) = I[\mathcal{M},\phi] = I_{\text{grav}}[\mathcal{M},g]+I_{\text{fields}}[\mathcal{M},\phi],
\end{equation}
where the total effective action $I[\mathcal{M},\phi]$ is evaluated on the $\mathcal{M},\phi$ for which it is minimised, subject to the appropriate boundary conditions. We call this minimal $\mathcal{M},\phi$ the `on-shell' configuration.

At leading order in the semiclassical approximation, we assume that the action is local, i.e.\ it may be written as 
\begin{equation}
    I[\mathcal{M},\phi] = \int_{\mathcal{M}}L[\phi],
\end{equation}
where the Lagrangian density $L[\phi]$ is a top form that depends locally on the fields $\phi$. Under a variation of the fields $\phi\to\phi+\delta\phi$, the change in the Lagrangian density may in general be written at linear order in $\delta\phi$ as
\begin{equation}
    \delta L[\phi] = E[\phi]\cdot\delta\phi + \dd(\theta[\phi,\delta\phi]),
\end{equation}
where $E[\phi]=0$ are the equations of motion, and $\cdot$ denotes a sum over all the fields. If we vary the fields from an on-shell configuration $\phi$ to an off-shell one $\phi+\delta\phi$, the action therefore changes by
\begin{equation}
    \delta I = \int_{\mathcal{N}}\theta[\phi,\delta\phi].
    \label{Equation: action variation}
\end{equation}
The consistency of the saddlepoint approximation requires the variational principle to be well defined. In other words, if the field variation $\delta\phi$ does not change the boundary conditions $\lambda$, then we must have $\delta I = 0$, so that $\phi$ genuinely extremises the action. This implies that the right hand side of~\eqref{Equation: action variation} can only depend on $\delta\phi$ through changes $\delta\lambda_1,\delta\lambda_2$ in the boundary conditions.

In the case $\lambda_1=\lambda_2$, the boundary conditions for the partition function have a $\mathbb{Z}_2$ symmetry under which they are complex conjugated, and reflected in time such that the past boundary is exchanged with the future boundary. Assuming this symmetry is not broken on-shell, it will fix a codimension 1 surface $\Sigma$ in $\mathcal{M}$. On this surface, all fields are invariant under the combination of time reversal and complex conjugation, and so can be Wick rotated to real Lorentzian fields. The Lorentzian development of those fields is what was depicted in Figure~\ref{Figure: set-up}. The black hole `lives' in this Lorentzian spacetime, even though we are using Euclidean partition functions to prepare the state.

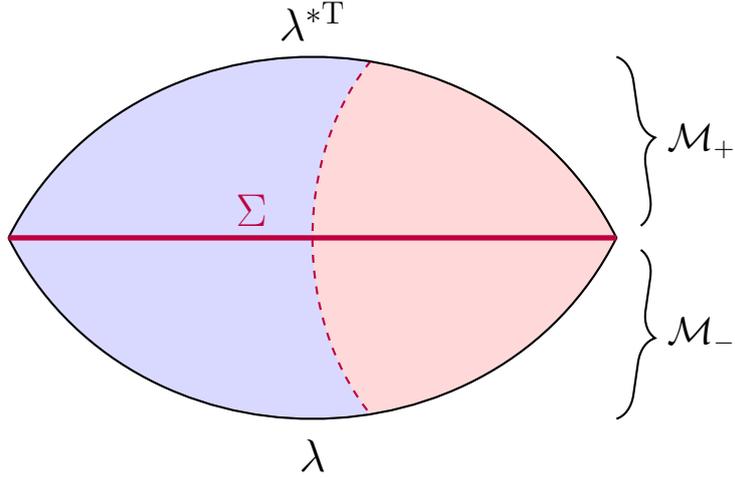
\begin{figure}
    \centering
    \begin{tikzpicture}[thick,scale=1.6]
        \fill[blue!15] (0,0) .. controls (1,-2) and (4,-2) .. (5,0)
            .. controls (4,2) and (1,2) .. (0,0);
        \begin{scope}
            \clip (0,0) .. controls (1,-2) and (4,-2) .. (5,0)
                .. controls (4,2) and (1,2) .. (0,0);
            \fill[red!15] (4,0) ellipse (1.5 and 2);
            \draw[purple,dashed] (4,0) ellipse (1.5 and 2);
        \end{scope}
        \draw (0,0) .. controls (1,-2) and (4,-2) .. (5,0)
            .. controls (4,2) and (1,2) .. (0,0);

        \node at (2.5,-1.8) {\Large$\lambda$};
        \node at (2.5,1.8) {\Large$\lambda^{*\mathrm{T}}$};
        \draw[line width=2pt,purple] (0,0) -- (5,0);
        \node[purple,above] at (2,0) {\Large$\Sigma$};
        \draw [decorate,decoration={brace,amplitude=10pt}] (5,1.5) -- (5.2,0.1);
        \draw [decorate,decoration={brace,amplitude=10pt}] (5.2,-0.1) -- (5,-1.5);
        \node at (5.7,0.8) {\large$\mathcal{M}_+$};
        \node at (5.7,-0.8) {\large$\mathcal{M}_-$};
    \end{tikzpicture}
    \caption{When $\lambda_1=\lambda_2$, all the fields are invariant under time reflection and complex conjugation. This symmetry fixes a surface $\Sigma$ which divides $\mathcal{M}$ into a past piece $\mathcal{M}_-$ and a future piece $\mathcal{M}_+$.}
    \label{Figure: same boundary conditions}
\end{figure}

The surface $\Sigma$ divides $\mathcal{M}$ into a past $\mathcal{M}_-$ and a future $\mathcal{M}_+$, such that $\partial\mathcal{M}_\pm=\Sigma\cup\mathcal{N}_{\pm}$, as shown in Figure~\ref{Figure: same boundary conditions}. If we fix some boundary conditions $\varphi$ for the fields on $\Sigma$, then we can write the on-shell action as
\begin{equation}
    I[\mathcal{M},\phi] = \min_{\varphi} \big(I[\mathcal{M}_-,\phi_-] + I[\mathcal{M}_+,\phi_+]\big),
    \label{Equation: action split}
\end{equation}
where $\phi_-,\phi_+$ are the fields in the past and future respectively, and
\begin{align}
    I[\mathcal{M}_-,\phi_-] &= \int_{\mathcal{M}_-}L[\phi_-] + \int_\Sigma D[\phi_-], \\
    I[\mathcal{M}_+,\phi_+] &= \int_{\mathcal{M}_+}L[\phi_+] - \int_\Sigma D[\phi_+]
\end{align}
are evaluated at their minima. Here, $D[\phi]$ only depends on the fields at $\Sigma$. The choice of $D[\phi]$ directly corresponds to the class of boundary conditions $\varphi$. To see this, suppose the past and future boundary conditions $\lambda$ are fixed. Then, under a variation of the fields from an on-shell configuration, we have
\begin{align}
    \delta I[\mathcal{M}_-,\phi_-] &= \int_\Sigma\big(\theta[\phi_-,\delta\phi_-] + \delta D[\phi_-]\big),\\
    \delta I[\mathcal{M}_+,\phi_+] &= -\int_\Sigma\big(\theta[\phi_+,\delta\phi_+] - \delta D[\phi_+]\big).
\end{align}
For the variational principle to be well-defined, the right hand sides above should vanish if we keep $\varphi$ fixed. Thus they can only depend on $\delta\phi_{\pm}$ via $\delta\varphi$. Different choices of $D[\phi]$ modify the dependence on $\delta\phi$, and so specify different classes of boundary conditions $\varphi$. Let us absorb $D[\phi]$ into the Lagrangian, i.e.\ redefine $L\to L+\dd{D}$.

In terms of the quantum theory, what we have just done amounts to the insertion of a resolution of the identity
\begin{equation}
    I = \int\Dd{\varphi}\ket{\varphi}\bra{\varphi}
\end{equation}
at $\Sigma$. The minimisation over $\varphi$ in~\eqref{Equation: action split} then comes from a saddlepoint approximation in the $\Dd{\varphi}$ integral. The states $\ket{\varphi}$ specify the boundary conditions at $\Sigma$. The time reflection and complex conjugation invariance of the fields at $\Sigma$ ensures that these states $\ket{\varphi}$ are genuine states in $\mathcal{H}$.

\subsection{Entropy from a replica trick}

\begin{figure}
    \centering
    \begin{subfigure}{0.4\textwidth}
        \centering
        \begin{tikzpicture}[scale=0.6,thick]
            \fill[blue!10] (0,0) .. controls (1,0) and (3,0) .. (4,0.2)
                .. controls (4.4,2.2) and (4.4,5.9) .. (4,7.9)
                .. controls (3.5,8) and (3.1,8) .. (3,8)
                -- (3,6) 
                .. controls (2.5,5.5) and (0.5,5.5) .. (0,5.5)
                -- (0,0);
            \fill[red!10] (4,0.2) .. controls (6,0.6) and (8,3) .. (9,4)
                .. controls (8,5) and (6,7.5) .. (4,7.9)
                .. controls (4.4,5.9) and (4.4,2.2) .. (4,0.2);
            \begin{scope}
                \clip (0,0) .. controls (1,0) and (3,0) .. (4,0.2)
                    .. controls (4.4,2.2) and (4.4,5.9) .. (4,7.9)
                    .. controls (3.5,8) and (3.1,8) .. (3,8)
                    -- (3,6) 
                    .. controls (2.5,5.5) and (0.5,5.5) .. (0,5.5)
                    -- (0,0);
                \fill[blue!25] (2.5,-1) .. controls (2.5,0.5) and (1.5,6) .. (0,6.5) -- (0,0);
            \end{scope}

            \draw (0,5.5) -- (0,0) .. controls (1,0) and (3,0) .. (4,0.2)
                .. controls (6,0.6) and (8,3) .. (9,4)
                .. controls (8,5) and (6,7.5) .. (4,7.9)
                .. controls (3.5,8) and (3.1,8) .. (3,8)
                -- (3,6);
            \begin{scope}
                \clip (4,0.2) .. controls (6,0.6) and (8,3) .. (9,4)
                    .. controls (8,5) and (6,7.5) .. (4,7.9)
                    .. controls (4.4,5.9) and (4.4,2.2) .. (4,0.2);
                \draw[line width=2pt,purple] (9,4) .. controls (8,3.2) and (6,3) .. (4,3);
            \end{scope}
            \node[purple] at (6.4,2.5) {\large$R$};
            \draw[very thick, blue!40!gray] (3,6) -- (0,3);
            \draw[decorate,decoration={snake,amplitude=0.8,segment length=3.5}] (3,6) .. controls (2.5,5.5) and (0.5,5.5) .. (0,5.5);
            \draw[dashed,purple] (4,0.2) .. controls (4.4,2.2) and (4.4,5.9) .. (4,7.9);

            \begin{scope}[-{Latex[]},orange!80!black]
                \draw (2.1,4) -- (5.1,7);
                \draw (2.1,3.7) -- (5.3,6.9);
                \draw (2.1,4.3) -- (4.9,7.1);
            \end{scope}
            \begin{scope}[-{Latex[]},blue!80!white]
                \draw (2,4) -- (0.8,5.2);
                \draw (2,3.7) -- (0.6,5.1);
                \draw (2,4.3) -- (1.0,5.3);
            \end{scope}
        \end{tikzpicture}
        \caption{\large$t<t_{\text{Page}}$}
    \end{subfigure}
    \begin{subfigure}{0.4\textwidth}
        \centering
        \begin{tikzpicture}[scale=0.6,thick]
            \fill[blue!10] (0,0) .. controls (1,0) and (3,0) .. (4,0.2)
                .. controls (4.4,2.2) and (4.4,5.9) .. (4,7.9)
                .. controls (3.5,8) and (3.1,8) .. (3,8)
                -- (3,6) 
                .. controls (2.5,5.5) and (0.5,5.5) .. (0,5.5)
                -- (0,0);
            \fill[red!10] (4,0.2) .. controls (6,0.6) and (8,3) .. (9,4)
                .. controls (8,5) and (6,7.5) .. (4,7.9)
                .. controls (4.4,5.9) and (4.4,2.2) .. (4,0.2);
            \begin{scope}
                \clip (0,0) .. controls (1,0) and (3,0) .. (4,0.2)
                    .. controls (4.4,2.2) and (4.4,5.9) .. (4,7.9)
                    .. controls (3.5,8) and (3.1,8) .. (3,8)
                    -- (3,6) 
                    .. controls (2.5,5.5) and (0.5,5.5) .. (0,5.5)
                    -- (0,0);
                \fill[blue!25] (2.5,-1) .. controls (2.5,0.5) and (1.5,6) .. (0,6.5) -- (0,0);
            \end{scope}

            \draw (0,5.5) -- (0,0) .. controls (1,0) and (3,0) .. (4,0.2)
                .. controls (6,0.6) and (8,3) .. (9,4)
                .. controls (8,5) and (6,7.5) .. (4,7.9)
                .. controls (3.5,8) and (3.1,8) .. (3,8)
                -- (3,6);
            \draw[very thick, blue!40!gray] (3,6) -- (0,3);
            \draw[decorate,decoration={snake,amplitude=0.8,segment length=3.5}] (3,6) .. controls (2.5,5.5) and (0.5,5.5) .. (0,5.5);
            \draw[dashed,purple] (4,0.2) .. controls (4.4,2.2) and (4.4,5.9) .. (4,7.9);

            \begin{scope}[-{Latex[]},orange!80!black]
                \draw (2.1,4) -- (5.1,7);
                \draw (2.1,3.7) -- (5.3,6.9);
                \draw (2.1,4.3) -- (4.9,7.1);
            \end{scope}
            \begin{scope}[-{Latex[]},blue!80!white]
                \draw (2,4) -- (0.8,5.2);
                \draw (2,3.7) -- (0.6,5.1);
                \draw (2,4.3) -- (1.0,5.3);
            \end{scope}

            \begin{scope}
                \clip (4,0.2) .. controls (6,0.6) and (8,3) .. (9,4)
                    .. controls (8,5) and (6,7.5) .. (4,7.9)
                    .. controls (4.4,5.9) and (4.4,2.2) .. (4,0.2);
                \draw[line width=2pt,purple] (9,4) .. controls (7,5.5) and (6,6) .. (4,6.5);
            \end{scope}
            \draw[line width=2pt, purple] (0,4.5) .. controls (1,4.5) and (1.7,5.1) .. (1.9,5.2);
            \fill[purple] (1.9,5.2) circle (0.1);
            \node[purple] at (6.4,4.9) {\large$R$};
            \node[purple] at (0.6,4.1) {\large$I$};
        \end{tikzpicture}
        \caption{\large$t>t_{\text{Page}}$}
    \end{subfigure}
    \caption{\mbox{\textbf{(a)}\ Before the Page time,} there is no island, and the entropy of the Hawking radiation is just given by the bulk entropy on the surface $R$. \mbox{\textbf{(b)}\ After the Page time,} enough entanglement has been built up by the Hawking process for an island to appear inside the black hole. Accounting for this in the island formula reproduces the Page curve.}
    \label{Figure: islands appearing}
\end{figure}
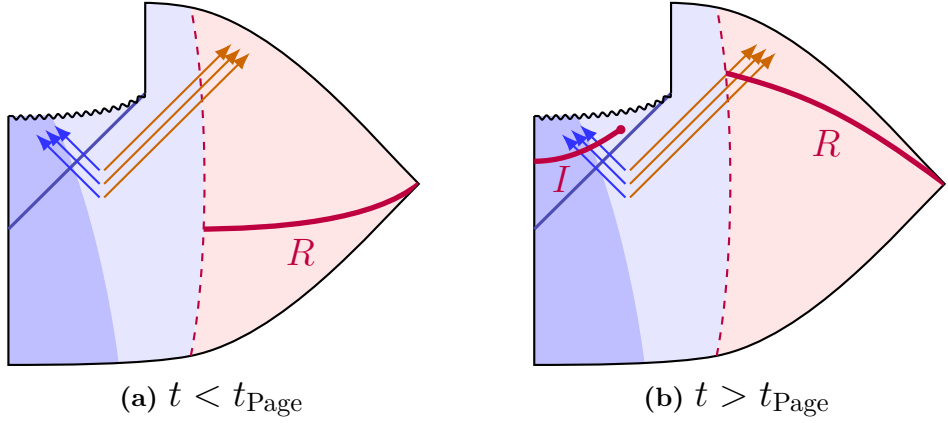

Given a state $\ket{\lambda}\in\mathcal{H}_G\otimes\mathcal{H}_R$, we can compute the reduced state $\rho_R$ of the Hawking radiation by taking a partial trace over $\mathcal{H}_G$. The entropy of the Hawking radiation is then given by the von Neumann entropy of $\rho_R$:
\begin{equation}
    S = -\tr(\rho_R(\lambda)\log\rho_R(\lambda)) \qq{where} \rho_R(\lambda) = \tr_G\ket{\lambda}\bra{\lambda}.
\end{equation}
According to~\cite{Flurry0,Flurry1,Flurry2,Flurry3,Flurry4,Flurry5}, $S$ is given by the island formula~\eqref{Equation: island formula}. At early times in the evaporation of a black hole, there is no island, but after the Page time an island appears in the black hole region, as shown in Figure~\ref{Figure: islands appearing}.

Let us now give an indication of where the island formula comes from.

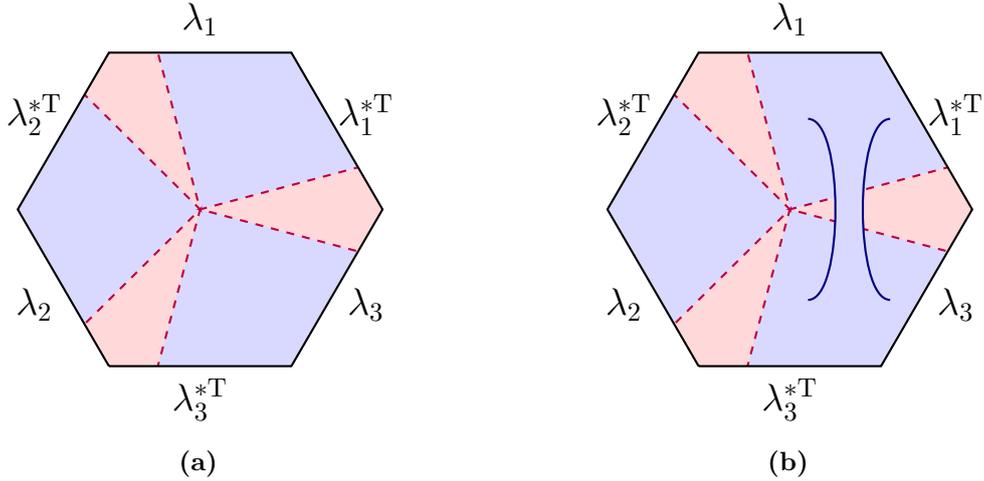
\begin{figure}
    \centering
    \begin{subfigure}{0.45\textwidth}
        \centering
        \begin{tikzpicture}[thick,scale=1.2]
            \fill[blue!15] (240:2) -- ++(0:2) -- ++(60:2) -- ++(120:2) -- ++(180:2) -- ++ (240:2) -- ++(300:2);
            \begin{scope}
                \clip (240:2) -- ++(0:2) -- ++(60:2) -- ++(120:2) -- ++(180:2) -- ++ (240:2) -- ++(300:2);
                \draw[purple,dashed,fill=red!15] (-15:3) -- (0,0) -- (15:3);
                \draw[purple,dashed,fill=red!15] (105:3) -- (0,0) -- (135:3);
                \draw[purple,dashed,fill=red!15] (225:3) -- (0,0) -- (255:3);
            \end{scope}
            \draw (240:2) -- ++(0:2) -- ++(60:2) -- ++(120:2) -- ++(180:2) -- ++ (240:2) -- ++(300:2);
            \node at (30:2.1) {\large$\lambda_1^{*\mathrm{T}}$};
            \node at (90:2.1) {\large$\lambda_1$};
            \node at (150:2.1) {\large$\lambda_2^{*\mathrm{T}}$};
            \node at (210:2.1) {\large$\lambda_2$};
            \node at (270:2.1) {\large$\lambda_3^{*\mathrm{T}}$};
            \node at (330:2.1) {\large$\lambda_3$};
        \end{tikzpicture}
        \caption{}
    \end{subfigure}
    \begin{subfigure}{0.45\textwidth}
        \centering
        \begin{tikzpicture}[thick,scale=1.2]
            \fill[blue!15] (240:2) -- ++(0:2) -- ++(60:2) -- ++(120:2) -- ++(180:2) -- ++ (240:2) -- ++(300:2);
            \begin{scope}
                \clip (240:2) -- ++(0:2) -- ++(60:2) -- ++(120:2) -- ++(180:2) -- ++ (240:2) -- ++(300:2);
                \draw[purple,dashed,fill=red!15] (-15:3) -- (0,0) -- (15:3);
                \draw[purple,dashed,fill=red!15] (105:3) -- (0,0) -- (135:3);
                \draw[purple,dashed,fill=red!15] (225:3) -- (0,0) -- (255:3);
            \end{scope}
            \draw (240:2) -- ++(0:2) -- ++(60:2) -- ++(120:2) -- ++(180:2) -- ++ (240:2) -- ++(300:2);
            \node at (30:2.1) {\large$\lambda_1^{*\mathrm{T}}$};
            \node at (90:2.1) {\large$\lambda_1$};
            \node at (150:2.1) {\large$\lambda_2^{*\mathrm{T}}$};
            \node at (210:2.1) {\large$\lambda_2$};
            \node at (270:2.1) {\large$\lambda_3^{*\mathrm{T}}$};
            \node at (330:2.1) {\large$\lambda_3$};
            \fill[blue!15] (0.2,-1) .. controls (0.6,-1) and (0.6,1) .. (0.2,1)
                -- (1.1,1) .. controls (0.7,1) and (0.7,-1) .. (1.1,-1);
            \draw[blue!50!black] (0.2,-1) .. controls (0.6,-1) and (0.6,1) .. (0.2,1);
            \draw[blue!50!black] (1.1,-1) .. controls (0.7,-1) and (0.7,1) .. (1.1,1);
        \end{tikzpicture}
        \caption{}
    \end{subfigure}
    \caption{The partition function $Z_{(n)}[\lambda_n,\dots,\lambda_1]$ should be computed on manifolds $\widetilde{\mathcal{M}}_n$ with the boundary conditions shown (the case shown is for $n=3$). The $n$ gravitational regions are disconnected in \textbf{(a)}, and this is what we would do in an ordinary QFT. However, since this is a theory of gravity, we should allow all topologies in the gravitational regions, including those which connect various copies with wormholes, for example like in \textbf{(b)}.}
    \label{Figure: replica partition}
\end{figure}

To compute the inner product $\braket{\lambda_2}{\lambda_1}$ we summed over degrees of freedom in both the asymptotic and gravitational regions -- colloquially, this is called `gluing'. On the other hand, the partial trace in $\rho_R(\lambda)=\tr_G\ket{\lambda}\bra{\lambda}$ involves just a sum over the fields in the gravitational region. The composition of several reduced states
\begin{equation}
    \rho_R(\lambda_n)\rho_R(\lambda_{n-1})\dots \rho_R(\lambda_2)\rho_R(\lambda_1)
\end{equation}
involves then gluing the future asymptotic region of $\rho_R(\lambda_i)$ to the past asymptotic region of $\rho_R(\lambda_{i+1})$ for $i=1,\dots,n-1$. By taking a trace, we then glue the future asymptotic region of $\rho_R(\lambda_n)$ back to the past asymptotic region of $\rho_R(\lambda_1)$, and the result may be written in terms of partition functions:
\begin{equation}
    \tr\big(\rho_R(\lambda_n)\dots \rho_R(\lambda_1)\big) = \frac{Z_{(n)}[\lambda_n,\dots,\lambda_1]}{Z_{(1)}[\lambda_n]\dots Z_{(1)}[\lambda_1]}.
    \label{Equation: partition replica}
\end{equation}
Here $Z_{(1)}[\lambda] = Z(\lambda,\lambda)$, and $Z_{(n)}[\lambda_n,\dots,\lambda_1]$ is the `replica' partition function for the boundary conditions just described. It may also be computed with a semiclassical approximation, using the effective gravitational action. We have
\begin{equation}
    -\log Z_{(n)}[\lambda_n,\dots,\lambda_1] = I_{\text{grav}}[\widetilde{\mathcal{M}}_n,g] - \log Z_{\text{QFT}}[\widetilde{\mathcal{M}}_n,g],
    \label{Equation: semiclassical gravity replica}
\end{equation}
where $\widetilde{\mathcal{M}}_n$ is a manifold with metric $g$ obeying the appropriate boundary conditions, and the right hand side is evaluated at the $\widetilde{\mathcal{M}}_n$ for which it is minimised. In the asymptotic region, the topology of the $n$ copies is fixed. However, in the gravitational region, we should allow $\widetilde{\mathcal{M}}_n$ to have any topology that fills in the boundary conditions. In particular, this means that there can be `replica wormholes' in the gravitational region that connect the $n$ copies. This is demonstrated in Figure~\ref{Figure: replica partition}.

Before we take the semiclassical limit for the QFT fields as well, we have to insert so-called twist operators $\mathcal{T}_n\dots\mathcal{T}_1$, which ensure that the various copies of the QFT are connected in the right way~\cite{Twist1,Twist2}. The semiclassical limit of $Z_{\text{QFT}}$ then includes the expectation value of the twist operators, as well as the effective action:
\begin{equation}
    -\log Z_{\text{QFT}}[\widetilde{\mathcal{M}}_n,g] = I_{\text{fields}}[\widetilde{\mathcal{M}}_n,\phi] - \log\expval{\mathcal{T}_n\dots\mathcal{T}_1}.
\end{equation}
Overall, in the full semiclassical limit, the replica partition function is given by
\begin{equation}
    -\log Z_{(n)}[\lambda_n,\dots,\lambda_1] = I[\widetilde{\mathcal{M}}_n,\phi] - \log\expval{\mathcal{T}_n\dots\mathcal{T}_1}.
\end{equation}

\begin{figure}
    \centering
    \begin{tikzpicture}[thick,scale=2.3]
        \begin{scope}[yscale=0.4]
            \fill[blue!15] (240:2) -- ++(0:2) -- ++(60:2) -- ++(120:2) -- ++(180:2) -- ++ (240:2) -- ++(300:2);
            \begin{scope}
                \clip (240:2) -- ++(0:2) -- ++(60:2) -- ++(120:2) -- ++(180:2) -- ++ (240:2) -- ++(300:2);
                \draw[purple,dashed,fill=red!15] (-15:3) -- (0,0) -- (15:3);
                \draw[purple,dashed,fill=red!15] (105:3) -- (0,0) -- (135:3);
                \draw[purple,dashed,fill=red!15] (225:3) -- (0,0) -- (255:3);
            \end{scope}
            \draw (240:2) -- ++(0:2) -- ++(60:2) -- ++(120:2) -- ++(180:2) -- ++ (240:2) -- ++(300:2);
            \node at (30:2.1) {\large$\lambda^{*\mathrm{T}}$};
            \node at (-0.2,2.1) {\large$\lambda$};
            \node at (150:2.1) {\large$\lambda^{*\mathrm{T}}$};
            \node at (210:2.1) {\large$\lambda$};
            \node at (270:2.1) {\large$\lambda^{*\mathrm{T}}$};
            \node at (330:2.1) {\large$\lambda$};

            \draw[-{Latex},very thick] (0,0)++(-30:2.2 and 3.1) arc (-30:-150:2.2 and 3.1);
            \node at (-1.6,-2.9) {\Large$\mathbb{Z}_n$};
        \end{scope}
        \fill[blue!15] (-1.5,0.2) .. controls (-1.1,0.2) and (-1.1,0.7) .. (-1.1,1.2)
            .. controls (-1.1,1.7) and (-0.7,2.5) .. (0,2.5) 
            .. controls (0.7,2.5) and (0.9,1.9) .. (0.9,1.4)
            .. controls (0.9,0.9) and (0.7,0.4) .. (1.1,0.4)
            -- (0.4,0.4) .. controls (0.8,0.4) and (0.6,1.3) .. (0.5,1.5)
            .. controls (0.3,1.9) and (-0.3,0) .. (1,-0.2)
            -- (0.3,-0.5) .. controls (0.6,-0.2) and (0,0) .. (0,1)
            .. controls (0,1.1) and (0,1.6) .. (-0.3,1.6) 
            .. controls (-0.7,1.6) and (-0.85,1.2) .. (-0.95,0.8)
            .. controls (-1,0.6) and (-1,0.2) .. (-0.8,0.2) -- cycle;

        \begin{scope}[blue!50!black]
            \draw (-1.5,0.2) .. controls (-1.1,0.2) and (-1.1,0.7) .. (-1.1,1.2)
                .. controls (-1.1,1.7) and (-0.7,2.5) .. (0,2.5) 
                .. controls (0.7,2.5) and (0.9,1.9) .. (0.9,1.4)
                .. controls (0.9,0.9) and (0.7,0.4) .. (1.1,0.4);
            \draw (0.4,0.4) .. controls (0.8,0.4) and (0.6,1.3) .. (0.5,1.5)
                .. controls (0.3,1.9) and (-0.3,0) .. (1,-0.2);
            \draw (0.3,-0.5) .. controls (0.6,-0.2) and (0,0) .. (0,1)
                .. controls (0,1.1) and (0,1.6) .. (-0.3,1.6) 
                .. controls (-0.7,1.6) and (-0.85,1.2) .. (-0.95,0.8)
                .. controls (-1,0.6) and (-1,0.2) .. (-0.8,0.2);
        \end{scope}

        \begin{scope}[purple!80]
            \draw (-0.24,1.6) .. controls (-0.29,1.71) and (-0.1,2.1) .. (0,2.2);
            \draw (0.39,1.55) .. controls (0.35,1.77) and (0.1,2.15) .. (0,2.2);
            \draw (0.05,2.5) .. controls (0.01,2.4) and (0.01,2.3) .. (0,2.2);
            \draw[dotted] (-0.24,1.6) -- (0.05,1.7);
            \draw[dotted] (0.39,1.55) -- (0.05,1.7);
            \draw[dotted] (0.05,2.5) .. controls (0.09,2.4) and (0.13,2.1) .. (0.05,1.7);
        \end{scope}

        \fill[purple] (0,2.2) circle (0.04);
        \fill[purple!40] (0.05,1.7) circle (0.04);
        \node[right] at (1.2,2.23) {\Large$\widetilde\Upsilon$};
        \draw[-{Latex},black!70,semithick] (1.2,2.23) .. controls (0.4,2.4) and (0.2,2.4) .. (0.03,2.23);
        \draw[-{Latex},black!70,semithick] (1.2,2.23) .. controls (0.4,2.4) and (0.25,1.9) .. (0.08,1.73);
        \node at (-0.35,1.8) {\large$I$};
        \node at (-0.1,2.35) {\large$I$};
        \node at (0.4,1.85) {\large$I$};
    \end{tikzpicture}
    \caption{The on-shell configuration for $Z_{(n)}=Z_{(n)}[\lambda,\dots,\lambda]$ has a $\mathbb{Z}_n$ replica symmetry. In general there will be a codimension 2 surface $\widetilde{\Upsilon}$ in the gravitational region which is fixed by $\mathbb{Z}_n$, here shown as two points. $\widetilde{\Upsilon}$ is the boundary of $n$ copies of a codimension 1 surface $I$, which are cyclically permuted by $\mathbb{Z}_n$.}
    \label{Figure: entropy replica symmetry}
\end{figure}
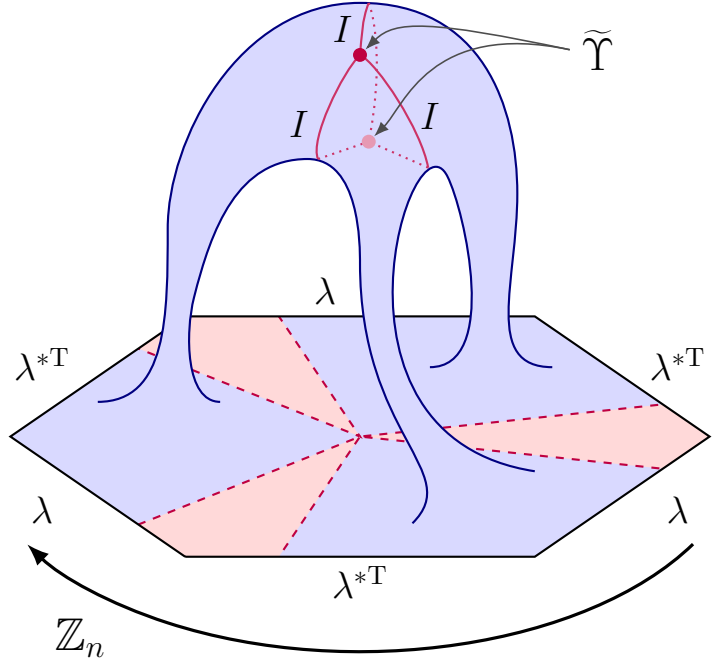

To compute the von Neumann entropy of $\rho_R=\rho_R(\lambda)$ we can use the `replica trick',
\begin{equation}
    S = -\tr(\rho_R\log\rho_R) = -\left.\partial_n\tr((\rho_R)^n)\right|_{n=1}.
\end{equation}
The aim is to compute $\tr((\rho_R)^n)$ for integer $n$ in terms of the replica partition function. Then we analytically continue in $n$ in order to compute the $n$ derivative. Assuming we have done this, we can write
\begin{equation}
    S = -\left.\partial_n\qty(\frac{\log Z_{(n)}}{n})\right|_{n=1},
    \label{Equation: replica trick partition}
\end{equation}
where $Z_{(n)}=Z_{(n)}[\lambda,\lambda,\dots,\lambda]$. The boundary conditions for this partition function have a $\mathbb{Z}_n$ replica symmetry which cyclically permutes the $n$ copies of the theory. Assuming this symmetry is not spontaneously broken, it will also be obeyed by the on-shell configuration $\widetilde{\mathcal{M}}_n,\phi$, as shown in Figure~\ref{Figure: entropy replica symmetry}. Then we can define the quotient manifold $\mathcal{M}_n=\widetilde{\mathcal{M}}_n/\mathbb{Z}_n$, and the field configuration $\phi$ reduces unambiguously to $\mathcal{M}_n$. In $\widetilde{\mathcal{M}}_n$ there is a codimension 2 surface $\widetilde{\Upsilon}$ which is fixed by $\mathbb{Z}_n$, and in $\mathcal{M}_n$ this becomes a surface $\Upsilon$ with a conical defect of opening angle $2\pi/n$. Then we can write
\begin{equation}
    I[\widetilde{\mathcal{M}}_n,\phi] = I[\widetilde{\mathcal{M}}_n\setminus\widetilde{\Upsilon},\phi] = nI[\mathcal{M}_n\setminus\Upsilon,\phi] = n\qty(I[\mathcal{M}_n,\phi] + (n-1)\frac{\operatorname{Area}(\Upsilon)}{4G}).
\end{equation}
The first equality follows from the assumption that the fields in $\widetilde{\mathcal{M}}_n$ are regular near $\widetilde{\Upsilon}$, while the second comes from replica symmetry. The area term on the right hand side cancels out any contributions that come from integrating over the conical defect $\Upsilon$ in the action.\footnote{This is the appropriate form for this term in general relativity. In higher derivative theories of gravity the area should be replaced by some other functional of the geometry of $\Upsilon$~\cite{HigherDerivative1,HigherDerivative2}.} We can quite easily analytically continue this action in $n$, because all the dependence on $n$ comes in the strength of the conical defect, which can be varied continuously.

\begin{figure}
    \centering
    \begin{tikzpicture}[thick,scale=1.6]
        \fill[blue!15] (0,0) .. controls (1,-2) and (4,-2) .. (5,0)
            .. controls (4,2) and (1,2) .. (0,0);
        \begin{scope}
            \clip (0,0) .. controls (1,-2) and (4,-2) .. (5,0)
                .. controls (4,2) and (1,2) .. (0,0);
            \fill[red!15] (4,0) ellipse (1.5 and 2);
            \draw[purple,dashed] (4,0) ellipse (1.5 and 2);
        \end{scope}
        \draw (0,0) .. controls (1,-2) and (4,-2) .. (5,0)
            .. controls (4,2) and (1,2) .. (0,0);

        \node at (2.5,-1.8) {\Large$\lambda$};
        \node at (2.5,1.8) {\Large$\lambda^{*\mathrm{T}}$};
        \draw[line width=2pt,purple] (5,0) -- (2.5,0);
        \draw[line width=2pt,purple] (1,0) -- (2,0);
        \fill[purple] (1,0) circle (0.05);
        \fill[purple] (2,0) circle (0.05);
        \node[above] at (1.5,0) {\Large$I$};
        \node[above] at (3.75,0) {\Large$R$};
        \draw[->] (1,0)++(30:0.18) arc (30:330:0.18);
        \draw[->] (2,0)++(150:0.18) arc (150:-150:0.18);
        \node at (0.7,0.25) {\large$\frac{2\pi}n$};
    \end{tikzpicture}
    \caption{After quotienting by the $\mathbb{Z}_n$ replica symmetry, we get a manifold $\mathcal{M}_n$ with a single copy of the boundary conditions. There is a conical defect at $\Upsilon=\partial I$, and we insert twist operators on $I\cup R$.}
    \label{Figure: replica quotiented}
\end{figure}
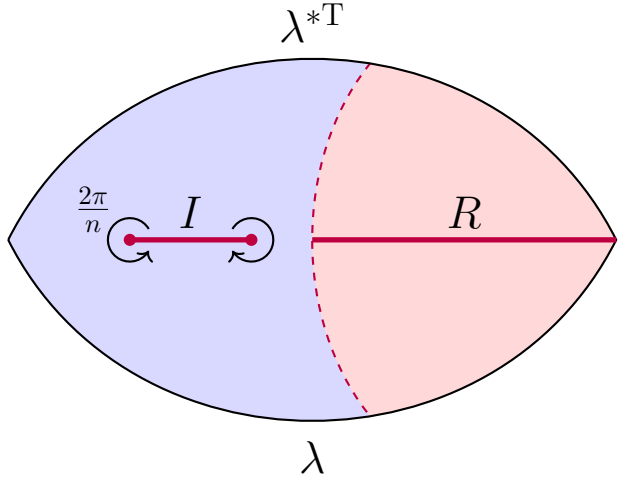

Note that the fixed surface $\widetilde{\Upsilon}$ is the boundary of $n$ copies of a region $I$ in $\widetilde{\mathcal{M}}_n$ which are cyclically permuted by replica symmetry. These become a single region $I$ in $\mathcal{M}_n$. This $I$ is the island. The overall structure of $\mathcal{M}_n$ is shown in Figure~\ref{Figure: replica quotiented}.

So the entropy may be written
\begin{equation}
    S = -\left.\partial_n\qty(I[\mathcal{M}_n,\phi] + (n-1)\frac{\operatorname{Area}(\Upsilon)}{4G} + \log\expval{\mathcal{T}_n\dots\mathcal{T}_1})\right|_{n=1}.
\end{equation}
Note that at $n=1$, there is no conical defect in $\mathcal{M}_n$. In fact, at $n=1$, $\mathcal{M}_n,\phi$ is the exact same on-shell configuration $\mathcal{M},\phi$ relevant to $\mathcal{Z}_{(1)}[\lambda]$. Computing the derivative in $n$ of $I[\mathcal{M}_n,\phi]$ at $n=1$ amounts to finding a linear perturbation of an on-shell action. But since the boundary conditions are fixed, by definition the action does not change at linear order, because we are on-shell. So the first term above vanishes. The second term is simple to compute. The third term is what we would get if we were computing an entropy using the replica trick in a ordinary perturbative QFT. The twist operators act on both the asymptotic region $R$ \emph{and} the island $I$. Thus, what the third term is computing is the bulk entropy of the combined region $I\cup R$. Combining all of this, we get
\begin{equation}
    S = \frac{\operatorname{Area}(\partial I)}{4G} + S_{\text{bulk}}(I\cup R).
\end{equation}
It remains to find the location of the island, and to do this we should note that $-\log Z_{(n)}$ should be minimised for all $n$. For $n$ close to 1, we can use the series expansion
\begin{equation}
    -\frac1{n}\log Z_{(n)} = I[\mathcal{M},\phi] + (n-1) S + \order{(n-1)^2}.
\end{equation}
We should minimise both terms appearing in this expansion individually. Minimising the first just means $\mathcal{M},\phi$ is on-shell. Note that we should take $n\ge1$, because this is the regime in which the replica trick works (according to Carlson's theorem). So minimising the second term is the same as minimising $S$. Therefore, we should compute the entropy $S$ using the island $I$ for which it is minimised, which is exactly the claim of the island formula.

\section{Uhlmann phase of Hawking radiation}
\label{Section: Uhlmann phase}

The quantity we now wish to compute is called the Uhlmann phase. This is not a widely known object, so let us briefly review how it is defined. For a more extensive description see~\cite{Uhlmann1992,Uhlmann1991,Uhlmann:sp,Braunstein,Jozsa:1994,Bures,Helstrom}, or for the holographic context see~\cite{Kirklin4}. The calculation that appears in the present work is essentially a more streamlined and generalised version of the one in that paper.

Given a curve of density matrices $\rho(t)$, $0\le t\le 1$ acting on a Hilbert space $\mathcal{H}$, we can always pick an auxiliary Hilbert space $\mathcal{H}'$, and find a curve of pure states $\ket{\psi(t)}$ in $\mathcal{H}\otimes\mathcal{H}'$ such that
\begin{equation}
    \rho(t) = \tr'\ket{\psi(t)}\bra{\psi(t)},
    \label{Equation: purifying}
\end{equation}
where $\tr'$ denotes a partial trace over $\mathcal{H}'$. Such states $\ket{\psi(t)}$ are said to be purifications of $\rho(t)$. If the curve also maximises the transition probability
\begin{equation}
    \abs{\braket{\psi(t+\epsilon)}{\psi(t)}}^2
\end{equation}
at leading order in small $\epsilon$, subject to \eqref{Equation: purifying}, then it is said to be `parallel'.

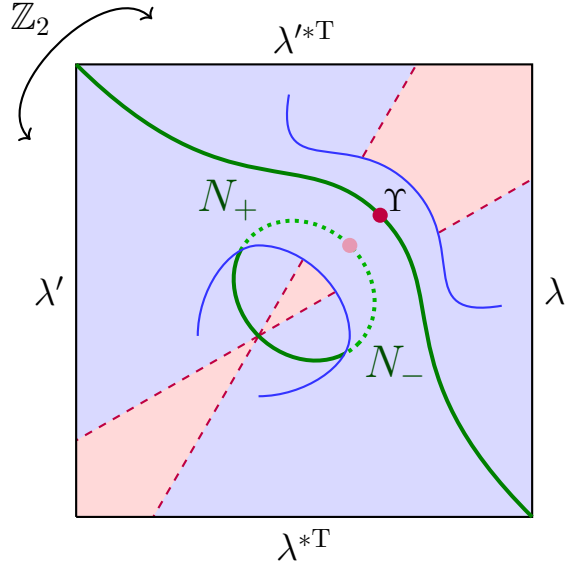
\begin{figure}
    \centering
    \begin{tikzpicture}[thick,scale=2]
        \fill[blue!15] (0,0) -- (3,0) -- (3,3) -- (0,3) -- (0,0);
        \begin{scope}
            \clip (0,0) -- (3,0) -- (3,3) -- (0,3) -- (0,0);
            \path (1.2,1.2) -- ++(30:5) coordinate (a);
            \path (1.2,1.2) -- ++(60:5) coordinate (b);
            \path (1.2,1.2) -- ++(240:5) coordinate (c);
            \path (1.2,1.2) -- ++(210:5) coordinate (d);
            \fill[red!15] (a) -- (d) -- (c) -- (b);
            \draw[purple,dashed] (1.2,1.2) -- (a);
            \draw[purple,dashed] (1.2,1.2) -- (b);
            \draw[purple,dashed] (1.2,1.2) -- (c);
            \draw[purple,dashed] (1.2,1.2) -- (d);
        \end{scope}
        \draw (0,0) -- (3,0) -- (3,3) -- (0,3) -- (0,0);

        \draw[shift={(1.5,1.5)},rotate=-45,green!50!black,line width=1.5pt] (0,0) ellipse (0.5 and {0.3*sqrt(2)});

        \fill[blue!15] (0.8,1.2) .. controls (0.8,1.5) and (1,1.8) .. (1.2,1.8)
            .. controls (1.5,1.8) and (1.8,1.5) .. (1.8,1.2)
            .. controls (1.8,1) and (1.5,0.8) .. (1.2,0.8)
            -- (2.8,1.4) .. controls (2.2,1.3) and (2.6,1.8) .. (2.2,2.2)
            .. controls (1.8,2.6) and (1.3,2.2) .. (1.4,2.8);
        \begin{scope}
            \clip (0.8,1.2) .. controls (0.8,1.5) and (1,1.8) .. (1.2,1.8)
                .. controls (1.5,1.8) and (1.8,1.5) .. (1.8,1.2)
                .. controls (1.8,1) and (1.5,0.8) .. (1.2,0.8)
                -- (2.8,1.4) .. controls (2.2,1.3) and (2.6,1.8) .. (2.2,2.2)
                .. controls (1.8,2.6) and (1.3,2.2) .. (1.4,2.8);
            \draw[shift={(1.5,1.5)},rotate=-45,green!70!black,line width=1.5pt,dotted] (0,0) ellipse (0.5 and {0.3*sqrt(2)});
        \end{scope}

        \draw[green!50!black,line width=1.5pt] (3,0) .. controls (2,1) and (2.5,1.5) .. (2,2)
            .. controls (1.5,2.5) and (1,2) .. (0,3);

        \draw[blue!80] (0.8,1.2) .. controls (0.8,1.5) and (1,1.8) .. (1.2,1.8)
            .. controls (1.5,1.8) and (1.8,1.5) .. (1.8,1.2)
            .. controls (1.8,1) and (1.5,0.8) .. (1.2,0.8);
        \draw[blue!80] (2.8,1.4) .. controls (2.2,1.3) and (2.6,1.8) .. (2.2,2.2)
            .. controls (1.8,2.6) and (1.3,2.2) .. (1.4,2.8);

        \fill[purple] (2,2) circle (0.05);
        \fill[purple!40] (1.8,1.8) circle (0.05);
        \node at (2.1,2.1) {$\Upsilon$};

        \node[above] at (1.5,3) {\large$\lambda^{\prime*\mathrm{T}}$};
        \node[below] at (1.5,0) {\large$\lambda^{*\mathrm{T}}$};
        \node[left] at (0,1.5) {\large$\lambda'$};
        \node[right] at (3,1.5) {\large$\lambda$};
        \node[green!30!black] at (1,2.1) {\Large $N_+$};
        \node[green!30!black] at (2.1,1) {\Large $N_-$};

        \draw[<->] (-0.3,2.5) .. controls (-0.6,2.8) and (0.2,3.6) .. (0.5,3.3);
        \node at (-0.3,3.3) {\large$\mathbb{Z}_2$};
    \end{tikzpicture}
    \caption{The configuration relevant for the partition function $Z_{(2k)}[\lambda',\lambda,\dots,\lambda',\lambda]$. By quotienting $\widetilde{\mathcal{M}}_{2k}$ by $\mathbb{Z}_k$, we get the manifold $\widehat{\mathcal{M}}_{2k}=\widetilde{\mathcal{M}}_{2k}/\mathbb{Z}_k$ shown here, with a conical defect at $\Upsilon$ of opening angle $2\pi/k$. This manifold, and the fields on it, enjoy a remaining $\mathbb{Z}_2$ symmetry that complex conjugates all the fields, and reflects them across the gravitational region in the way shown. This symmetry fixes a surface $N_+\cup N_-$, which goes through $\Upsilon$ and meets the boundary of the asymptotic region.}
    \label{Figure: Zk quotient}
\end{figure}

If $\rho(t)$ is a closed curve of invertible density matrices, purified by $\ket{\psi(t)}$ in a parallel manner,\footnote{Note that in general the curve of pure states $\ket{\psi(t)}$ need not be closed, even if the curve of mixed states $\rho(t)$ is.} then the Uhlmann phase $\gamma$ of $\rho(t)$ is defined via
\begin{equation}
    e^{i\gamma} = \lim_{m\to\infty}\braket{\psi(0)}{\psi(1)}\braket*{\psi(1)}{\psi\qty(\frac{m-1}{m})}\dots\braket*{\psi\qty(\frac2m)}{\psi\qty(\frac1m)}\braket*{\psi\qty(\frac1m)}{\psi(0)}.
\end{equation}
For any closed curve $\rho(t)$ there are many possible choices for the parallel purifying curve $\ket{\psi(t)}$, but it may be shown that $\gamma$ does not depend on this choice. 

Suppose $\lambda(t)$ is a closed curve of boundary conditions, and let $\rho_R(t) = \tr_G\ket{\lambda(t)}\bra{\lambda(t)}$ be the corresponding Hawking radiation states. Typical density matrices of subsystems in (regularised) quantum field theories and quantum gravity are invertible, and we will assume this is true for $\rho_R$. To compute the Uhlmann phase of $\rho_R(t)$, we need to find a parallel purifying curve $\ket{\psi(t)}$. We will use Uhlmann's theorem~\cite{Jozsa:1994}, which says that two purifications $\ket{\psi},\ket{\psi'}$ of density matrices $\rho,\rho'$ have maximal transition probability if and only if
\begin{equation}
    \abs{\braket{\psi'}{\psi}} = \tr(\sqrt{\sqrt{\rho}\rho'\sqrt{\rho}}).
    \label{Equation: Uhlmann theorem}
\end{equation}
The right hand side is known as the fidelity of $\rho$ and $\rho'$. 

\subsection{Fidelity and parallel states from a replica trick} 
To employ Uhlmann's theorem, we will first need to compute the fidelity of two Hawking radiation states
\begin{equation}
    \rho_R = \tr_G\ket{\lambda}\bra{\lambda} \qq{and} \rho_R' = \tr_G\ket{\lambda'}\bra{\lambda'}.
\end{equation}
We can do this with a replica trick. First we compute
\begin{equation}
    \tr(\qty(\sqrt{\sqrt{\rho_R}\rho_R'\sqrt{\rho_R}})^{2k}) = \tr((\rho_R\rho_R')^k)
    \label{Equation: fidelity replica}
\end{equation}
for integer $k$, and then we analytically continue to $k=\frac12$. 

In terms of partition functions, we have
\begin{equation}
    \tr((\rho_R\rho_R')^k) = \frac{Z_{(2k)}[\lambda,\lambda',\dots,\lambda,\lambda']}{Z_{(1)}[\lambda]^kZ_{(1)}[\lambda']^k}.
    \label{Equation: replicated fidelity}
\end{equation}
In the partition function in the numerator, the boundary conditions are alternately given by $\lambda$ and $\lambda'$ on the $2k$ copies. We may write this in terms of the usual semiclassical approximation,
\begin{equation}
    -\log Z_{(2k)}[\lambda,\lambda',\dots,\lambda,\lambda'] = I[\widetilde{\mathcal{M}}_{2k},\phi] - \log \expval{\mathcal{T}_{2k}\dots\mathcal{T}_1}.
    \label{Equation: replicated fidelity partition}
\end{equation}
When $\lambda\ne \lambda'$, the previous $\mathbb{Z}_{2k}$ replica symmetry of the boundary conditions is explicitly broken down to a $\mathbb{Z}_k$ subgroup. Quotienting by this remaining symmetry, we can write
\begin{equation}
    I[\widetilde{\mathcal{M}}_{2k},\phi] = I[\widetilde{\mathcal{M}}_{2k}\setminus\widetilde{\Upsilon},\phi] = I[\widehat{\mathcal{M}}_{2k}\setminus\Upsilon,\phi],
\end{equation}
where $\widehat{\mathcal{M}}_{2k} = \widetilde{\mathcal{M}}_{2k}/\mathbb{Z}_k$, and $\Upsilon$ is the image of the $\mathbb{Z}_k$ fixed codimension 2 surface $\widetilde{\Upsilon}$. There is a conical defect of opening angle $2\pi/k$ at $\Upsilon$.

\begin{figure}
    \centering
    \begin{tikzpicture}[thick,scale=1.2]
        \begin{scope}[shift={(13,0)},xscale=-1]
            \fill[blue!15] (0,0) .. controls (1,2) and (3.5,2) .. (6,1)
                .. controls (5.5,0.5) and (5,0.2) .. (4.5,0)
                .. controls (5,-0.2) and (5.5,-0.5) .. (6,-1)
                .. controls (3.5,-2) and (1,-2) .. (0,0);
            \begin{scope}
                \clip (0,0) .. controls (1,2) and (3.5,2) .. (6,1)
                    .. controls (5.5,0.5) and (5,0.2) .. (4.5,0)
                    .. controls (5,-0.2) and (5.5,-0.5) .. (6,-1)
                    .. controls (3.5,-2) and (1,-2) .. (0,0);
                \draw[purple,dashed,fill=red!15] (0,0) ellipse (2 and 3);
            \end{scope}
            \fill[white]
                (3.5,0) .. controls (3,0.5) and (2.5,0.5) .. (2,0)
                .. controls (2.5,-0.5) and (3,-0.5) .. (3.5,0);

            \draw (0,0) .. controls (1,2) and (3.5,2) .. (6,1);
            \draw (0,0) .. controls (1,-2) and (3.5,-2) .. (6,-1);
            \draw[green!50!black,line width=1.5pt] 
                (6,1) .. controls (5.5,0.5) and (5,0.2) .. (4.5,0)
                .. controls (5,-0.2) and (5.5,-0.5) .. (6,-1);
            \draw[green!50!black,line width=1.5pt]
                (3.5,0) .. controls (3,0.5) and (2.5,0.5) .. (2,0)
                .. controls (2.5,-0.5) and (3,-0.5) .. (3.5,0);
            \fill[purple] (4.5,0) circle (0.08);
            \fill[purple] (3.5,0) circle (0.08);
            \node at (4,.8) {\large$\Upsilon$};
            \draw[->] (3.5,0)++(110:0.18) arc (110:-110:0.18);
            \draw[->] (4.5,0)++(35:0.18) arc (35:325:0.18);
            \node at (4,-0.1) {\large$\frac{\pi}k$};
            \node[green!30!black] at (5.3,0.8) {\large$N_+$};
            \node[green!30!black] at (5.3,-0.8) {\large$N_-$};
            \node[green!30!black] at (2.75,0.65) {\large$N_+$};
            \node[green!30!black] at (2.75,-0.65) {\large$N_-$};
            \draw[-{Latex},black!70,semithick] (3.85,0.65) -- (3.6,0.2);
            \draw[-{Latex},black!70,semithick] (4.15,0.65) -- (4.4,0.2);
            \node at (3,2) {\Large$\lambda^{\prime*\mathrm{T}}$};
            \node at (3,-2) {\Large$\lambda$};
        \end{scope}
        \fill[blue!15] (0,0) .. controls (1,2) and (3.5,2) .. (6,1)
            .. controls (5.5,0.5) and (5,0.2) .. (4.5,0)
            .. controls (5,-0.2) and (5.5,-0.5) .. (6,-1)
            .. controls (3.5,-2) and (1,-2) .. (0,0);
        \begin{scope}
            \clip (0,0) .. controls (1,2) and (3.5,2) .. (6,1)
                .. controls (5.5,0.5) and (5,0.2) .. (4.5,0)
                .. controls (5,-0.2) and (5.5,-0.5) .. (6,-1)
                .. controls (3.5,-2) and (1,-2) .. (0,0);
            \draw[purple,dashed,fill=red!15] (0,0) ellipse (2 and 3);
        \end{scope}
        \fill[white]
            (3.5,0) .. controls (3,0.5) and (2.5,0.5) .. (2,0)
            .. controls (2.5,-0.5) and (3,-0.5) .. (3.5,0);

        \draw (0,0) .. controls (1,2) and (3.5,2) .. (6,1);
        \draw (0,0) .. controls (1,-2) and (3.5,-2) .. (6,-1);
        \draw[green!50!black,line width=1.5pt] 
            (6,1) .. controls (5.5,0.5) and (5,0.2) .. (4.5,0)
            .. controls (5,-0.2) and (5.5,-0.5) .. (6,-1);
        \draw[green!50!black,line width=1.5pt]
            (3.5,0) .. controls (3,0.5) and (2.5,0.5) .. (2,0)
            .. controls (2.5,-0.5) and (3,-0.5) .. (3.5,0);
        \fill[purple] (4.5,0) circle (0.08);
        \fill[purple] (3.5,0) circle (0.08);
        \node at (4,.8) {\large$\Upsilon$};
        \draw[->] (3.5,0)++(110:0.18) arc (110:-110:0.18);
        \draw[->] (4.5,0)++(35:0.18) arc (35:325:0.18);
        \node at (4,-0.1) {\large$\frac{\pi}k$};
        \node[green!30!black] at (5.3,0.8) {\large$N_+$};
        \node[green!30!black] at (5.3,-0.8) {\large$N_-$};
        \node[green!30!black] at (2.75,0.65) {\large$N_+$};
        \node[green!30!black] at (2.75,-0.65) {\large$N_-$};
        \draw[-{Latex},black!70,semithick] (3.85,0.65) -- (3.6,0.2);
        \draw[-{Latex},black!70,semithick] (4.15,0.65) -- (4.4,0.2);
        \node at (3,2) {\Large$\lambda'$};
        \node at (3,-2) {\Large$\lambda^{*\mathrm{T}}$};

        \draw [decorate,decoration={brace,amplitude=10pt}] (0,2.5) -- (6.2,2.5);
        \node at (3,3.2) {\large$\mathcal{M}_{2k}^*$};
        \draw [decorate,decoration={brace,amplitude=10pt}] (6.8,2.5) -- (13,2.5);
        \node at (10,3.2) {\large$\mathcal{M}_{2k}$};
    \end{tikzpicture}
    \caption{By cutting $\widehat{\mathcal{M}}_{2k}$ along $N_+\cup N_-$, we get two manifolds $\mathcal{M}_{2k}$ and $\mathcal{M}_{2k}^*$, which are related to each other by reflection and complex conjugation.}
    \label{Figure: hat M split}
\end{figure}
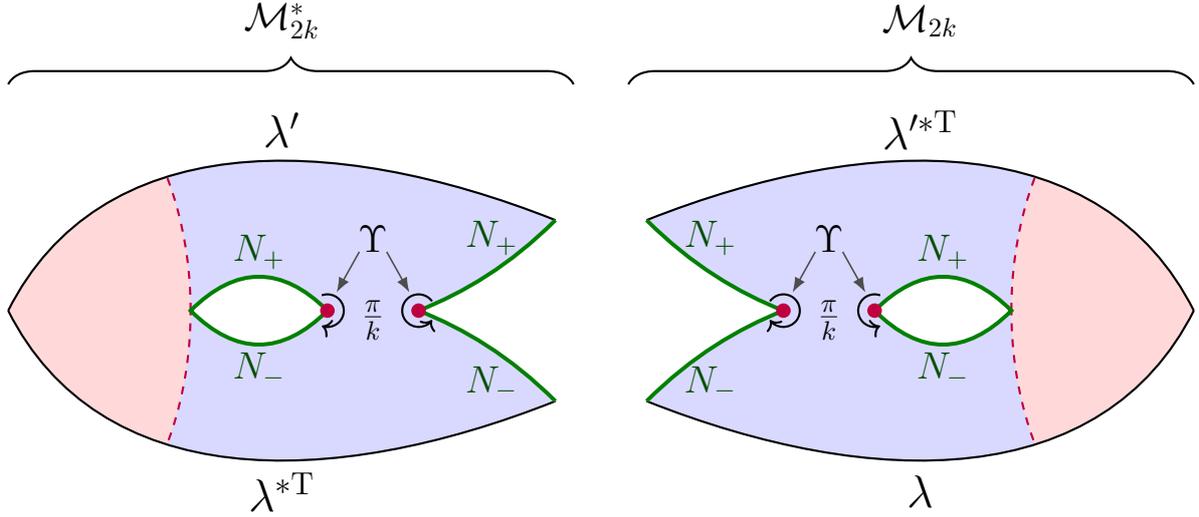

There is a remaining $\mathbb{Z}_2$ symmetry of the boundary conditions for $\widehat{\mathcal{M}}_{2k},\phi$. This symmetry complex conjugates the boundary conditions, and reflects them across the remaining two copies of the gravitational region. Assuming this symmetry is not broken, it will fix a codimension 1 surface $N_+\cup N^-$ which divides $\widehat{\mathcal{M}}_{2k}$ into two pieces, as depicted in Figure~\ref{Figure: Zk quotient}. We will call the two pieces $\mathcal{M}_{2k}$ and $\mathcal{M}^*_{2k}$; they are shown separately in Figure~\ref{Figure: hat M split}. The two pieces $N_+$ and $N_-$ of $N_+\cup N_-$ meet at $\Upsilon$ and the boundary of the asymptotic region. $N_+$ denotes the piece closest to the boundary conditions $\lambda'$, and $N_-$ denotes the piece closest to the boundary conditions $\lambda$. We can then write
\begin{equation}
    I[\widehat{\mathcal{M}}_{2k}\setminus\Upsilon,\phi] = I[\mathcal{M}_{2k}\setminus\Upsilon,\phi] + I[\mathcal{M}^*_{2k}\setminus\Upsilon,\phi] = 2\operatorname{Re}(I[\mathcal{M}_{2k}\setminus\Upsilon,\phi]),
\end{equation}
where the last line follows from the $\mathbb{Z}_2$ symmetry. Note that, in $\mathcal{M}_{2k}$, the angle around $\Upsilon$ from $N_-$ to $N_+$ is $\pi/k$, because the $\mathbb{Z}_2$ symmetry divides the $2\pi/k$ conical defect in two. Accounting for this with an area term, we have
\begin{equation}
    I[\mathcal{M}_{2k}\setminus\Upsilon,\phi] = I[\mathcal{M}_{2k},\phi] + (2k-1)\frac{\operatorname{Area}[\Upsilon]}{4G}.
\end{equation}

Because the fields at $N_+$ and $N_-$ are invariant under the $\mathbb{Z}_2$, they define states $\ket{\varphi_+}$ and $\ket{\varphi_-}$. These states depend on $k$. Let us now analytically continue to $k=\frac12$, so that $\mathcal{M}_{2k}\to\mathcal{M}_1$, which has no conical defect at $\Upsilon$. Let $U$ be a unitary operator that maps $\ket{\varphi_-}$ to $\ket{\varphi_+}$.  Note that $U$ only changes the fields in the gravitational region, so it is a unitary operator acting on $\mathcal{H}_G$. We can think of $I[\mathcal{M}_1,\phi]$ as the effective action in the presence of an insertion of $U$ on a surface $N$. $N_+$ lies just to the `future' of $N$, while $N_-$ lies just to the `past' of $N$. To be more precise, consider the correlator
\begin{equation}
    \mel{\lambda'}{U\otimes I}{\lambda} = \frac{Z^U(\lambda',\lambda)}{\sqrt{Z(\lambda,\lambda)Z(\lambda',\lambda')}},
    \label{Equation: U partition}
\end{equation}
where $I$ is the identity acting on $\mathcal{H}_R$, and $Z^U$ is the partition function with $U$ inserted. We can compute this with an effective action
\begin{equation}
    -\log Z^U(\lambda',\lambda) = I^U[\mathcal{M},\phi].
\end{equation}
The right hand side is the effective action for boundary conditions $\lambda'$ in the future, and $\lambda$ in the past. Additionally, the fields in this action are acted on by $U$ at $N$. We must compute this action for the configuration $\mathcal{M},\phi$ which minimises it. But note that these boundary conditions, and the action of $U$, are exactly the same conditions that apply to $I[\mathcal{M}_1,\phi]$. Thus, we can write
\begin{equation}
    -\log Z^U(\lambda',\lambda) = I[\mathcal{M}_1,\phi].
    \label{Equation: U partition 2}
\end{equation}

The contribution of the twist operators in~\eqref{Equation: replicated fidelity partition} is a symmetric function of $\lambda$ and $\lambda'$,
\begin{equation}
    F(\lambda,\lambda') = -\frac1{2k}\log\expval{\mathcal{T}_{2k}\dots\mathcal{T}_1}.
\end{equation}
We know that this function is real, because the effective action is real, and so is~\eqref{Equation: fidelity replica} (since it is the trace of a Hermitian operator). From the calculations of the previous section, we also know that at $\lambda=\lambda'$ we can expand this function near $2k=1$ in terms of the bulk entropy for those boundary conditions:
\begin{equation}
    F(\lambda,\lambda) = (2k-1)S_{\text{bulk}}(I\cup R)|_{\lambda} + \order{(2k-1)^2} = \order{2k-1}.
\end{equation}
(We know that the $\order{1}$ piece vanishes because at $k=1$ we have $\expval{\mathcal{T}_{2k}\dots\mathcal{T}_1}=\expval{1}=1$.) At this point, let us assume that $\lambda'$ is only a small perturbation of $\lambda$. Assuming that $F(\lambda,\lambda')$ is sufficiently smooth near $\lambda=\lambda'$, for a small enough perturbation we can bound $F(\lambda,\lambda')$ with either
\begin{equation}
    F(\lambda,\lambda) \le F(\lambda,\lambda') \le F(\lambda',\lambda')
\end{equation}
or
\begin{equation}
    F(\lambda,\lambda) \ge F(\lambda,\lambda') \ge F(\lambda',\lambda')
\end{equation}
But since in both cases the upper and lower bounds are $\order{2k-1}$, we must have
\begin{equation}
    F(\lambda,\lambda') = \order{2k-1}.
\end{equation}
Therefore, when we analytically continue to $k=\frac12$ we will get
\begin{equation}
    -\log\expval{\mathcal{T}_{2k}\dots\mathcal{T}_1}\big|_{k=\frac12} = 0.
\end{equation}

Using~\eqref{Equation: replicated fidelity} and continuing to $k=\frac12$, we therefore get the formula for the fidelity when $\lambda'$ is a small perturbation of $\lambda$:
\begin{equation}
    \tr(\sqrt{\sqrt{\rho_R}\rho_R'\sqrt{\rho_R}}) = \frac1{\sqrt{Z_{(1)}[\lambda]Z_{(1)}[\lambda']}} \exp\qty(-\operatorname{Re}(I[\mathcal{M}_1,\phi])).
    \label{Equation: hawking fidelity}
\end{equation}
The location of $\Upsilon$ is constrained as before by considering the behaviour of the partition function for small $2k-1$. From this point on, let us write $\lambda' = \lambda+\epsilon\delta\lambda$, where $\abs{\epsilon}\ll1$. We have
\begin{equation}
    -\frac1{2k}\log Z_{(2k)}[\lambda,\lambda+\epsilon\delta\lambda,\dots] = \big(I[\mathcal{M},\bar\phi]+\order{\epsilon}\big) + (2k-1) \big(S+\order{\epsilon}\big) + \order{(2k-1)^2},
\end{equation}
where $\bar\phi$ is the field configuration at $\epsilon=0$. We will need to minimise this for arbitray small $\epsilon$. This means we need to minimise each of the terms in this expansion, so, up to $\order{\epsilon}$ corrections, $\Upsilon$ is located at the boundary of the island $I$. 

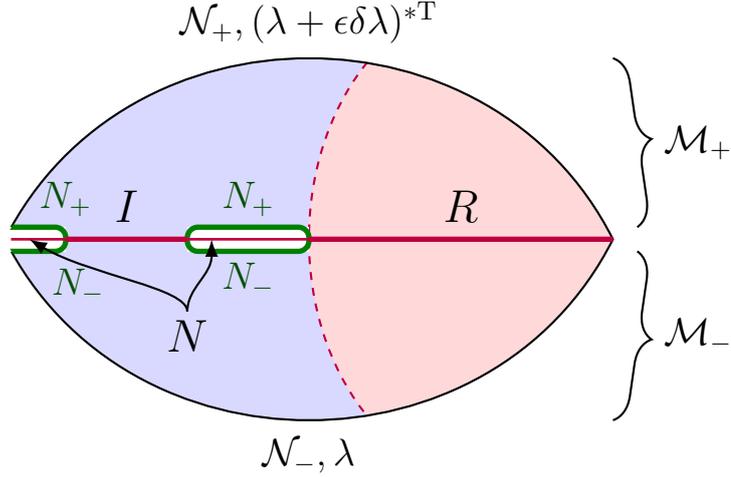
\begin{figure}
    \centering
    \begin{tikzpicture}[thick,scale=1.6]
        \fill[blue!15] (0,0) .. controls (1,-2) and (4,-2) .. (5,0)
            .. controls (4,2) and (1,2) .. (0,0);
        \fill[white] (-1,-0.1) rectangle (0.5,0.1);
        \begin{scope}
            \clip (0,0) .. controls (1,-2) and (4,-2) .. (5,0)
                .. controls (4,2) and (1,2) .. (0,0);
            \fill[red!15] (4,0) ellipse (1.5 and 2);
            \draw[purple,dashed] (4,0) ellipse (1.5 and 2);
        \end{scope}
        \fill[white] (1.5,-0.1) rectangle (2.5,0.1);
        \draw[rounded corners,line width=2pt,green!50!black] (1.5,-0.1) rectangle (2.5,0.1);
        \draw (0,0) .. controls (1,-2) and (4,-2) .. (5,0)
            .. controls (4,2) and (1,2) .. (0,0);
        \fill[white] (-1,-0.1) rectangle (0.5,0.1);
        \begin{scope}
            \clip (0,0) .. controls (1,-2) and (4,-2) .. (5,0)
                .. controls (4,2) and (1,2) .. (0,0);
            \draw[rounded corners,line width=2pt,green!50!black] (-1,-0.1) rectangle (0.5,0.1);
        \end{scope}

        \node at (2.5,-1.8) {\large$\mathcal{N}_-,\lambda$};
        \node at (2.5,1.8) {\large$\mathcal{N}_+,(\lambda+\epsilon\delta\lambda)^{*\mathrm{T}}$};
        \draw[line width=1pt,purple] (0.05,0) -- (5,0);
        \draw[line width=2pt,purple] (0.5,0) -- (1.5,0);
        \draw[line width=2pt,purple] (2.5,0) -- (5,0);
        \node[above] at (1,0.05) {\Large$I$};
        \node[above] at (3.75,0.05) {\Large$R$};
        \node[above,green!30!black] at (2,0.1) {\large$N_+$};
        \node[below,green!30!black] at (2,-0.1) {\large$N_-$};
        \node[above,green!30!black] at (0.5,0.1) {\large$N_+$};
        \node[below,green!30!black] at (0.6,-0.15) {\large$N_-$};
        \draw [decorate,decoration={brace,amplitude=10pt}] (5,1.5) -- (5.2,0.1);
        \draw [decorate,decoration={brace,amplitude=10pt}] (5.2,-0.1) -- (5,-1.5);
        \node at (5.7,0.8) {\large$\mathcal{M}_+$};
        \node at (5.7,-0.8) {\large$\mathcal{M}_-$};
        \node at (1.5,-0.8) {\Large$N$};
        \draw[-{Latex},black] (1.5,-0.6) .. controls (1.5,-0.4) and (1.7,-0.3) .. (1.7,0);
        \draw[-{Latex},black] (1.5,-0.6) .. controls (1.5,-0.4) and (0.7,-0.3) .. (0.2,0);
    \end{tikzpicture}
    \caption{At small $\epsilon$, computing the fidelity of $\rho_R$ and $\rho_R'$ is equivalent to computing the absolute value of the inner product of two states $\ket{\lambda}$ and $\ket{\lambda'}$, with the insertion of a unitary operator $U$ acting at $N$. We have $N\cup I \cup R = \Sigma$, where $\Sigma$ was shown in Figure~\ref{Figure: same boundary conditions}.}
    \label{Figure: U insertion}
\end{figure}

At this point we can immediately write down a pair of states that satisfy Uhlmann's theorem for $\rho_R$ and $\rho_R'$. Suppose $V$ is any unitary operator acting on $\mathcal{H}_G$, and let
\begin{equation}
    \ket{\psi} = (V\otimes I)\ket{\lambda}, \qquad \ket{\psi'} = (VU^\dagger\otimes I)\ket{\lambda+\epsilon\delta\lambda},
\end{equation}
where $U$ was defined above. Clearly these states are purifications of $\rho_R,\rho_R'$ respectively. Also, we have
\begin{equation}
    \braket{\psi'}{\psi} = \mel{\lambda+\epsilon\delta\lambda}{U\otimes I}{\lambda}=\frac1{\sqrt{Z_{(1)}[\lambda]Z_{(1)}[\lambda+\epsilon\delta\lambda]}} \exp\qty(-I[\mathcal{M}_1,\phi]),
    \label{Equation: psi'psi overlap}
\end{equation}
where the second equality follows from~\eqref{Equation: U partition} and~\eqref{Equation: U partition 2}. Taking the absolute value of this, and using~\eqref{Equation: hawking fidelity}, we observe that
\begin{equation}
    \abs{\braket{\psi'}{\psi}} = \tr(\sqrt{\sqrt{\rho_R}\rho_R'\sqrt{\rho_R}}),
\end{equation}
as desired. This is illustrated in Figure~\ref{Figure: U insertion}. 

\subsection*{Overlap of parallel states}

Let us now consider the behaviour of the fields $\phi$ for $\lambda'=\lambda+\epsilon\delta\lambda$, for small $\epsilon$. At $\epsilon=0$, there is an additional $\mathbb{Z}_2$ symmetry on $\mathcal{M}_1$ that complex conjugates the fields, and reflects everything in time. At $\epsilon=0$, this symmetry implies that the fields on $N_-$ are equal to those on $N_+$. This implies that the unitary operator $U$ at $\epsilon=0$ is just the identity. When $\epsilon$ is small but non-zero, this $\mathbb{Z}_2$ symmetry is broken, but we can still use it to analyse the resulting perturbation to the fields at linear order. We will write
\begin{equation}
    \phi = \bar\phi+\frac12\epsilon\big(\delta\phi+\tilde\delta\phi\big),
\end{equation}
where $\bar\phi$ is the background field configuration at $\epsilon=0$, and under the broken $\mathbb{Z}_2$ symmetry we have
\begin{equation}
    \delta\phi \to \delta\phi,\qquad \tilde\delta\phi \to -\tilde\delta\phi.
    \label{Equation: broken Z2 variations}
\end{equation}
The variation $\delta\phi$ corresponds to a variation in the boundary conditions of $\delta\lambda$ at $\mathcal{N}_-$ and $\delta\lambda^{*\mathrm{T}}$ at $\mathcal{N}_+$. It is continuous as we go from $N_-$ to $N_+$. The other variation $\tilde\delta\phi$ corresponds to a variation in the boundary conditions of $-\delta\lambda$ at $\mathcal{N}_-$ and $\delta\lambda^{*\mathrm{T}}$ at $\mathcal{N}_+$. It is discontinuous as we go from $N_-$ to $N_+$. In particular it changes sign.

Note that $\bar\phi$ is the dominant field configuration for the partition function $Z_{(1)}[\lambda]$, and $\bar\phi+\epsilon\delta\phi$ is the dominant field configuration for the partition function $Z_{(1)}[\lambda+\epsilon\delta\lambda]$. Thus, we may write~\eqref{Equation: psi'psi overlap} as
\begin{align}
    \braket{\psi'}{\psi} &= \exp(\frac12 I[\mathcal{M},\bar\phi]+\frac12I[\mathcal{M},\bar\phi+\epsilon\delta\phi] - I[\mathcal{M}_1,\bar\phi+\frac12\epsilon(\delta\phi+\tilde\delta\phi)]+\order{\epsilon^2}) \\
                         &= \exp(-\frac12\epsilon\tilde\delta I[\mathcal{M}_1,\bar\phi]+\order{\epsilon^2}).
\end{align}
Here, $\epsilon \tilde{\delta} I$ denotes the variation of the action due to $\bar\phi\to \bar\phi+\epsilon\tilde{\delta}\phi$. In terms of $\theta[\bar\phi,\tilde\delta\phi]$, this can be written
\begin{align}
    \frac12\tilde\delta I[\mathcal{M}_1,\bar\phi] &= \frac12\qty(\int_{\partial\mathcal{M}_1}\theta[\bar\phi,\tilde\delta\phi])\\
                                                  &= \frac12\qty(\int_{\mathcal{N}_-}\theta[\bar\phi,\tilde\delta\phi] + \int_{\mathcal{N}_+}\theta[\bar\phi,\tilde\delta\phi] + \int_{N_-}\theta[\bar\phi,\tilde\delta\phi]+\int_{N_+}\theta[\bar\phi,\tilde\delta\phi])\\
                                         &= \int_{\mathcal{N}_+}\theta[\bar\phi,\tilde\delta\phi] + \int_{N_+}\theta[\bar\phi,\tilde\delta\phi].
\end{align}
The manifold $\mathcal{M}_1$ is shown in Figure~\ref{Figure: U insertion}, from which we can see that $\partial\mathcal{M}_1=\mathcal{N}_+\cup\mathcal{N}_-\cup N_+\cup N_-$, and so the second line holds. The third line follows from~\eqref{Equation: broken Z2 variations}. The change in boundary conditions for $\tilde\delta\phi$ and $\delta\phi$ are the same at $\mathcal{N}_+$, and $\theta|_{\mathcal{N}_+}$ only depends on the change in boundary conditions, so we can write
\begin{equation}
    \int_{\mathcal{N}_+}\theta[\bar\phi,\tilde\delta\phi] = \int_{\mathcal{N}_+}\theta[\bar\phi,\delta\phi].
\end{equation}
Also, it will be convenient to use a basis for the states on $N$ such that
\begin{equation}
    \int_{N_+}\theta[\bar\phi,\tilde\delta\phi-\delta\phi] = 0.
\end{equation}
As explained in the previous section, this can be done by adding appropriate boundary terms to the action. Then we have
\begin{equation}
    \braket{\psi'}{\psi} = \exp(-\epsilon\int_{\partial\mathcal{M}^+\cup N} \theta[\bar\phi,\delta\phi]+\order{\epsilon^2}).
    \label{Equation: overlap theta}
\end{equation}

\subsection{Uhlmann phase}
At this point, we are ready to compute the Uhlmann phase of a curve $\rho_R(t)$ of density matrices corresponding to a curve of boundary conditions $\lambda(t)$, $0\le t \le 1$. At large $m$, for each pair of density matrices $\rho(i/m)$ and $\rho_R((i+1)/m)$, we can use the derivation above to obtain a unitary operator $U_i$ such that
\begin{equation}
    \abs{\mel{\lambda((i+1)/m)}{U_i\otimes I}{\lambda(i/m)}} = \tr(\sqrt{\sqrt{\rho_R(i/m)}\rho_R((i+1)/m)\sqrt{\rho_R(i/m)}}).
\end{equation}
Each of the operators $U_i$ maps a state $\ket*{\varphi^i_-}$ on $N_-$ to a state $\ket*{\varphi^i_+}$ on $N_+$. These states come from the boundary conditions at $N_\pm$ corresponding to a field configuration $\phi_i = \bar\phi_i + \frac1{2m}\qty(\delta_i\phi+\tilde\delta_i\phi)$, where $\bar\phi_i$ is the dominant field configuration relevant to $Z_{(1)}[\lambda(i/m)]$, and 
\begin{equation}
    \bar\phi_{i+1}-\bar\phi_i = \frac1m\delta_i\phi + \order{\frac1{m^2}}.
\end{equation}
To be precise, we have
\begin{equation}
    \varphi_\pm^i = \left.\qty(\bar\phi_i+\frac1{2m}\big(\delta_i\phi+\tilde\delta_i\phi\big))\right|_{\text{b.c.s at }N_\pm}.
\end{equation}
By adding the appropriate boundary terms to the action, we can ensure that the states on $N_\pm$ are in a basis such that
\begin{equation}
    \int_{N_+}\theta[\bar\phi_i,\tilde\delta_i\phi-\delta_i\phi] = 0,
\end{equation}
or equivalently
\begin{equation}
    \int_{N_-}\theta[\bar\phi_i,\tilde\delta_i\phi+\delta_i\phi] = 0.
\end{equation}
It is possible to find a boundary term that enforces this condition for all $i$, because in field space the vector $(\tilde\delta_i\phi-\delta_i\phi)|_{N_+}$ is never tangent to the curve $\bar\phi(t)$. Since this condition determines the class of boundary conditions $\varphi$, we can use it to write
\begin{align}
    \varphi_+^i &= \left.\qty(\bar\phi_i+\frac1{m}\delta_i\phi)\right|_{\text{b.c.s at }N}, \\
    \varphi_-^i &= \bar\phi_i\big|_{\text{b.c.s at }N}.
    \label{Equation: varphi- barphi}
\end{align}
Therefore, $\ket*{\varphi^i_+}=\ket*{\varphi^{i+1}_-}$.

Then we define the sequence of states
\begin{align}
    \ket{\psi(0)} &= \ket{\lambda(0)},\\
    \ket{\psi(1/m)} &= (U_0^\dagger\otimes I)\ket{\lambda(1/m)},\\
                    &\dots \\
    \ket{\psi(i/m)} &= (U_0^\dagger U_1^\dagger \dots U_{i-1}^\dagger\otimes I)\ket{\lambda(i/m)},\\ 
                    &\dots \\
    \ket{\psi(1)} &= (U_0^\dagger U_1^\dagger \dots U_{m-1}^\dagger\otimes I)\ket{\lambda(1)}.
\end{align}
These states purify $\rho(0),\rho(1/n),\dots$, and satisfy~\eqref{Equation: Uhlmann theorem}, so as $m\to\infty$ we get a parallel purifying curve $\ket{\psi(t)}$.

To find the Uhlmann phase we need to compute the overlaps of these states. First, from~\eqref{Equation: overlap theta} we know that
\begin{equation}
    \braket{\psi((i+1)/m)}{\psi(i/m)} = \exp(-\frac1{m}\int_{\mathcal{N}_+\cup N}\theta[\bar\phi_i,\delta_i\phi]+\order{\frac1{m^2}}).
\end{equation}
We also need to compute 
\begin{equation}
    \braket{\psi(0)}{\psi(1)} = \mel{\lambda(0)}{U_0^\dagger U_1^\dagger \dots U_{m-1}^\dagger}{\lambda(0)}.
\end{equation}
But note that
\begin{align}
    \bra*{\varphi_0^-} U_0^\dagger U_1^\dagger \dots U_{m-1}^\dagger &= \bra*{\varphi_0^+} U_1^\dagger \dots U_{m-1}^\dagger \\
                                                                     &= \bra*{\varphi_1^-} U_1^\dagger \dots U_{m-1}^\dagger\\
                                                                     &= \dots \\
                                                                     &= \bra*{\varphi_m^-} = \bra*{\varphi_0^-},
\end{align}
where the last equality follows because the curve of states is closed. So $U_0^\dagger U_1^\dagger \dots U_{m-1}^\dagger$ maps $\ket*{\varphi^-_0}$ to itself. $\varphi^-_0$ are just the boundary conditions for $\bar\phi_0$ at $N$, so inserting this operator in between $\bra{\lambda(0)}$ and $\ket{\lambda(0)}$ actually has no effect at the semiclassical level, because the dominant field configuration is unchanged. Thus, we just have
\begin{equation}
    \mel{\lambda(0)}{U_0^\dagger U_1^\dagger \dots U_{m-1}^\dagger}{\lambda(0)} = \exp(\order{\frac1{m^2}}).
\end{equation}

Finally, the Uhlmann phase is given by
\begin{align}
    e^{i\gamma} &= \lim_{m\to\infty}\braket{\psi(0)}{\psi(1)}\braket*{\psi(1)}{\psi\qty(\frac{m-1}{m})}\dots\braket*{\psi\qty(\frac2m)}{\psi\qty(\frac1m)}\braket*{\psi\qty(\frac1m)}{\psi(0)}\\
                &= \lim_{m\to\infty}\prod_{i=0}^{m-1}\exp(-\frac1m \int_{\mathcal{N}^+\cup N}\theta(\bar\phi_i,\delta_i\phi) + \order{\frac1{m^2}}) \\
                &= \lim_{m\to\infty}\exp(-\frac1m\sum_{i=0}^{m-1}\int_{\mathcal{N}^+\cup N}\theta(\bar\phi_i,\delta_i\phi) + \order{\frac1{m^2}}).
\end{align}
Taking the $m\to\infty$ limit converts the sum to an integral, and we are left with
\begin{equation}
    \gamma = \int_C \Theta,
    \label{Equation: gamma C Theta}
\end{equation}
where $C$ is the curve of field configurations $\bar\phi(t)=\bar\phi(\lambda(t))$ relevant to the partition function $Z_{(1)}[\lambda(t)]$, and $\Theta$ is the one-form on field space whose component in the direction $\delta\phi$ at a field configuration $\phi$ is given by
\begin{equation}
    \Theta[\phi,\delta\phi] = i\int_{\mathcal{N}_+\cup N} \theta[\phi,\delta\phi].
    \label{Equation: Hawking Theta}
\end{equation}

\section{Classical information recovery}
\label{Section: Information recovery}

We now wish to show how to use the Uhlmann phase of the Hawking radiation to \changed[reconstruct the classical phase space of]{recover classical information contained in} the black hole interior. The main claim is that the field space 1-form $\Theta$ defined in~\eqref{Equation: Hawking Theta} is in fact a symplectic potential for $I\cup R$, i.e.\ the island and the radiation. This means its exterior derivative is equal to the symplectic form for the fields in $I\cup R$. To justify this claim, let us take the exterior derivative:
\begin{equation}
    \delta\Theta = i\int_{\mathcal{N}_+\cup N} \omega.
\end{equation}
Here, $\omega=\delta\theta$\changed{, and $\delta$ denotes an exterior derivative on field space}. Note that $\dd{\omega} = \delta(\delta L) = 0$ on-shell. All field configurations we are considering are on-shell. Thus we can use Stokes' theorem to write
\begin{equation}
    \delta\Theta = -i\int_{I\cup R} \omega,
\end{equation}
because $\mathcal{N}_+\cup N \cup I \cup R = \partial\mathcal{M}_+$, as can be verified from Figure~\ref{Figure: U insertion}. The above expression is in terms of Euclidean fields. The fields at $I\cup R$ are invariant under time reflection and complex conjugation, so we can Wick rotate to real Lorentzian fields. After doing so, the factor of $i$ goes away because there are an odd number of time derivatives, and we end up with
\begin{equation}
    \delta\Theta = \int_{I\cup R}\omega_{\text{Lorentzian}} = \Omega_{I\cup R}.
    \label{Equation: Theta I R}
\end{equation}
The right hand side here is exactly the definition of the symplectic form of $I\cup R$, according to the covariant phase space formalism~\cite{CVP1,CVP2,CVP3,CVP4,CVP5,CVP6,CVP7,CVP8,CVP9,CVP10,CVP11,CVP12,CVP13,CVP14,CVP15,CVP16}. So we can conclude that $\Theta$ is indeed a symplectic potential for $I\cup R$. Therefore, the curvature of the Uhlmann phase \changed{of the }Hawking radiation is the symplectic form of $I\cup R$.

\subsection{General information recovery}

Before explaining how to apply our results to the recovery of classical information from black holes, it is important to first describe what we mean by information recovery. There are various ways to formulate this -- we will use one that is reminiscent of error correction, and which is directly applicable to the black hole case.

We start with a system whose initial state takes values in some space $\mathcal{P}_1$. We also have another space $\mathcal{I}$, each element of which describes the state of some information in the system. To embed this information in the physical system we assume the existence of a map
\begin{equation}
    f:\quad\mathcal{I}\to\mathcal{P}_1, \quad i \mapsto x_i.
\end{equation}
For each possible value taken by the information $i\in\mathcal{I}$, $f(i)=x_i\in\mathcal{P}_1$ is the physical state containing that information. We assume that there exists some map $R_1:\mathcal{P}_1\to\mathcal{I}$ obeying $R_1(x_i)=i$. This means that, after embedding the information in the system, we can easily get it back using $R_1$.

However, we are interested in a process which could potentially erase the information. In particular, we will assume the existence of a space of final states $\mathcal{P}_2$, and a map $\mathscr{S}:\mathcal{P}_1\to\mathcal{P}_2$ which represents the evolution of the system. After this evolution, the state of the system is $\bar{x}_i = \mathscr{S}(x_i)$. The information is recoverable if there exists a map $R_2:\mathcal{P}_2\to \mathcal{I}$ which obeys $R_2(\bar{x}_i) = i$, because we can use this map to read the information from the final state.

The typical example is that of a diary on fire. The initial space $\mathcal{P}_1$ consists of all of the possible initial states of the diary, while each $i\in I$ specifies the diary's contents, i.e.\ the state $x_i$ has the property that $i$ is written in the diary. We then set the diary on fire and let the fire burn until it has exhausted its fuel -- this burning happens via the map $\mathscr{S}:\mathcal{P}_1\to\mathcal{P}_2$. The space $\mathcal{P}_2$ consists of all of the possible final states of the radiation emitted from the fire, as well as the leftover ash. In principle, this is quantum mechanically a unitary process, so the map $\mathscr{S}$ is invertible, and we can just write $R_2=R_1\circ\mathscr{S}^{-1}$. Thus, the information (i.e.\ the contents of the diary) is recoverable. On the other hand, burning is a highly chaotic process, and in practice we can't measure the final state of the ash and radiation with perfect accuracy. To account for this we can modify $\mathscr{S}$ so that it implements a coarse graining of the final state, and redefine $\mathcal{P}_2$ to contain coarse grained states. This renders $\mathscr{S}$ uninvertible -- so we can no longer use the previous trick to recover information. Indeed, information will typically be lost in such a process.

For what follows it is useful to have:
\begin{lemma}
    Information is recoverable if and only if, given any function $F:\mathcal{P}_1\to\RR$, one can construct a function $\bar{F}:\mathcal{P}_2\to \RR$ obeying
        $F(x_i)-F(x_j) = \bar{F}(\bar{x}_i)-\bar{F}(\bar{x}_j)$
    for all $i,j\in\mathcal{I}$.
\end{lemma}
\begin{proof}
    Suppose information is recoverable. Then we can set $\bar{F} = F\circ f \circ R_2$, and for all $i\in\mathcal{I}$ we have
    \begin{equation}
        \bar{F}(\bar{x}_i) = F(f(R_2(\bar{x}_i))) = F(f(i)) = F(x_i).
    \end{equation}
    Thus $F(x_i)-F(x_j) = \bar{F}(\bar{x}_i)-\bar{F}(\bar{x}_j)$ as required.

    For the opposite direction, let us pick a set of functions $a_n:\mathcal{I} \to \RR$, $n=1,2,\dots$ such that the map $A:i \mapsto (a_1(i),a_2(i),\dots)$ is injective (so that knowing the values of $a_1,a_2,\dots$ fully determines the state of the information). Moreover, let us fix a $j\in \mathcal{I}$, and pick $a_n$ such that $a_n(j)=0$ for all $n$. Let $F_n=a_n\circ R_1$; then by assumption there exists an $\bar{F}_n:\mathcal{P}_2\to \RR$ obeying
    \begin{equation}
        \bar{F}_n(\bar{x}_i) - \bar{F}_n(\bar{x}_j) = F_n(x_i)-\underbrace{F_n(x_j)}_{=a_n(j)=0} = a_n(i).
    \end{equation}
    Note that we can set $\bar{F}_n(\bar{x}_j) = 0$ by redefining $\bar{F}_n(\bar{x})\to \bar{F}_n(\bar{x})-\bar{F}_n(\bar{x}_j)$. Then we just have $\bar{F}_n(\bar{x}_i) = a_n(i)$, so by measuring $\bar{F}_n$ in the final state we get the values of $a_n$ for all $n$, and thus we can determine $i$. This yields the map\footnote{Note this map is only guaranteed to be defined on $\bar{x}=\bar{x}_i$ for $i\in\mathcal{I}$. But for other values of $\bar{x}$ we can just set $R_2(\bar{x})$ to be anything at all, since its behaviour on such final states is irrelevant to the question of information recovery.}
    \begin{equation}
        R_2 : \quad \bar{x} \mapsto A^{-1}(\bar{F}_1(\bar{x}),\bar{F}_2(\bar{x}),\dots),
    \end{equation}
    which obeys $R_2(\bar{x}_i)=i$ and so renders the information recoverable.
\end{proof}

\subsection{Black hole classical information recovery}

Let us now see how we can use the Uhlmann phase to recover classical information from the interior of a black hole. In particular, we will show how one can recover any function of the classical information by measuring the Uhlmann phase around an appropriate curve of states.

Let $\mathcal{P}_1=\mathcal{P}_{\text{classical}}$ be a phase space of all classical field configurations containing a black hole. Let us suppose for simplicity that after a certain time $T$ the classical field configuration outside of the black hole is the same for all configurations in $\mathcal{P}_{\text{classical}}$. This means that an observer who can only read the classical values of the fields outside the black hole at time $T$ has no way of knowing which state in $\mathcal{P}_{\text{classical}}$ is the true one. To model the information recovery capabilities of such an observer, we would set $\mathcal{P}_2$ to contain the possible classical field configurations outside of the black hole at time $T$, and $\mathscr{S}$ would be the restriction map from the total field configuration to this exterior field configuration. Since $\mathcal{P}_2$ only contains one configuration, and $\mathscr{S}$ maps everything to that configuration, the observer is incapable of recovering any non-trivial information from the black hole.

Let us therefore give our observer an upgrade: we now will also allow them access to the quantum state $\rho$ of the Hawking radiation it emits. If Hawking's original calculations were valid, this state would have no knowledge of the black hole interior, and so the observer would be no more capable of recovering information than they were before their upgrade. However, the recent breakthroughs~\cite{Flurry0,Flurry1,Flurry2,Flurry3,Flurry4,Flurry5} show that, past the Page time, the state of the radiation \emph{does} contain information about the interior.

For the observer with access to the Hawking radiation, $\mathcal{P}_2=\mathcal{P}_{\text{Hawking}}$ is the space of possible states of Hawking radiation at time $T$ arising from classical field configurations in $\mathcal{P}_{\text{classical}}$. Given a field configuration $\phi\in\mathcal{P}_{\text{initial}}$, we get a unique Hawking radiation state $\rho\in\mathcal{P}_{\text{Hawking}}$; let this define the map 
\begin{equation}
    \mathscr{S}:\quad \mathcal{P}_{\text{classical}}\to\mathcal{P}_{\text{Hawking}}, \quad \phi\mapsto \rho.
\end{equation}

As in the previous section, we can imagine that the field configuration $\phi$ contains a diary. We can write some classical information $i\in I$ into that diary; let the corresponding field configuration be denoted $\phi_i$, and let the corresponding Hawking radiation be denoted $\rho_i=\mathscr{S}(\phi_i)$. Let us take $T$ to be greater than the Page time, so that an island has formed inside of the black hole. We will assume that the diary falls into the black hole and is at time contained $T$ within the island\footnote{Otherwise it would not possible to recover the information in the diary. Eventually the island spans the entire black hole interior, so if the diary is not yet in the island we can just increase $T$ until it is.}. We will further assume that the fields away from the diary are (to a good approximation) independent of $i$ -- in other words, the information is localised to the diary.

We will show that the information in the diary is recoverable by using the lemma from the previous subsection. Suppose $F:\mathcal{P}_{\text{classical}}\to\RR$ is a function of the classical field configuration. Note that if we restrict to $\phi=\phi_i$, then by our assumptions $F(\phi_i)$ only depends on the fields in the vicinity of the diary. Thus, we can define a new function $\tilde{F}:\mathcal{P}_{\text{classical}}\to\RR$ such that
\begin{equation}
    \tilde{F}(\phi_i)= F(\phi_i),
\end{equation}
and such that $\tilde{F}(\phi)$ only depends on the fields in the vicinity of the diary for all $\phi\in\mathcal{P}_{\text{classical}}$.

Since $\mathcal{P}_{\text{classical}}$ is a phase space, it comes equipped with a symplectic structure $\Omega$, which according to the covariant phase space formalism can be written as the integral of the previously mentioned spacetime 2-form $\omega_{\text{Lorentzian}}$ over any Cauchy surface $\Sigma$:
\begin{equation}
    \Omega = \int_\Sigma\omega_{\text{Lorentzian}}.
\end{equation}
It is convenient to pick a $\Sigma$ which contains the island $I$. Let $V$ be the vector field on $\mathcal{P}_{\text{classical}}$ obeying
\begin{equation}
    \iota_V\Omega = -\delta{\tilde{F}},
    \label{Equation: tilde F hamiltonian}
\end{equation}
i.e.\ $V$ is the Hamiltonian vector field associated with $\tilde{F}$. Let $\delta_V\phi$ denote the variation of the field configuration associated with this vector field. Note that since $\tilde{F}$ only depends on the fields in the vicinity of the diary, $\delta_V\phi$ must vanish away from the diary, which means we can write
\begin{equation}
    \iota_V\Omega = \iota_V\Omega_{I\cup R},
    \label{Equation: V vector I R}
\end{equation}
where $\Omega_{I\cup R}$ was defined in~\eqref{Equation: Theta I R}.

Now let $\bar{V}=\mathscr{S}_* V$ be the pushforward of $V$ through the dynamical map $\mathscr{S}$. We can use $\bar{V}$ and the Uhlmann phase to construct a certain function $\bar{F}:\mathcal{P}_{\text{Hawking}}\to \RR$, in the following way. Let us fix a $\rho_0\in\mathcal{P}_{\text{Hawking}}$. Suppose $\rho\in \mathcal{P}_{\text{Hawking}}$ is connected to $\rho_0$ by a curve $D_1\subset\mathscr{S}$. Let $\bar\psi_t$ be the flow generated by the vector field $\bar{V}$, let
\begin{equation}
    D_2 = \{\bar\psi_t(\rho):t\in [0,\epsilon]\}, \quad D_3 = \bar\psi_\epsilon(D_1), \quad D_4 = \{\bar\psi_t(\rho_0):t\in [0,\epsilon]\},
\end{equation}
where $0<\epsilon\ll 1$, and let $D=D_1\cup D_2\cup D_3 \cup D_4$ be the closed curve obtained by taking the union of these four curves. Finally, let 
\begin{equation}
    \bar{F}(\rho) = \lim_{\epsilon\to 0}\frac{\gamma(D)}{\epsilon},
\end{equation}
where $\gamma(D)$ is the Uhlmann phase around the curve $D$.

\begin{figure}
    \centering
    \begin{tikzpicture}[scale=0.65]
        \begin{scope}[shift={(11,0)}]
            \draw[rounded corners,fill=black!5] (-3,-1.5) rectangle (5,8);
            \draw[very thick]
                (-0.5,0) coordinate    (M0) .. controls (2,2) and (2.1,4) ..
                coordinate[pos=0.1] (M1)
                coordinate[pos=0.2] (M2)
                coordinate[pos=0.3] (M3)
                coordinate[pos=0.4] (M4)
                coordinate[pos=0.5] (M5)
                coordinate[pos=0.6] (M6)
                coordinate[pos=0.7] (M7)
                coordinate[pos=0.8] (M8)
                coordinate[pos=0.9] (M9)
                (3,6) coordinate    (M10)
                node[midway,below right] {$D_1$};
            \draw[very thick, shift={(-1,1)}] 
                (0,0) coordinate    (N0) .. controls (2,1) and (2,3) ..
                coordinate[pos=0.1] (N1)
                coordinate[pos=0.2] (N2)
                coordinate[pos=0.3] (N3)
                coordinate[pos=0.4] (N4)
                coordinate[pos=0.5] (N5)
                coordinate[pos=0.6] (N6)
                coordinate[pos=0.7] (N7)
                coordinate[pos=0.8] (N8)
                coordinate[pos=0.9] (N9)
                (3,6) coordinate    (N10)
                node[midway,above left] {$D_3$};

            \draw[very thick] (M0) -- (N0) node[midway,below left] {$D_4$};
            \draw[-{latex}] (M1) -- (N1);
            \draw[-{latex}] (M2) -- (N2);
            \draw[-{latex}] (M3) -- (N3);
            \draw[-{latex}] (M4) -- (N4);
            \draw[-{latex}] (M5) -- (N5);
            \draw[-{latex}] (M6) -- (N6);
            \draw[-{latex}] (M7) -- (N7);
            \draw[-{latex}] (M8) -- (N8)node[midway, above] {\footnotesize$\bar V$};
            \draw[-{latex}] (M9) -- (N9);
            \draw[very thick] (M10) -- (N10) node[midway, above right] {$D_2$};

            \fill (-0.5,0) circle (0.15) node[below] {\footnotesize$\rho_0$};
            \fill (3,6) circle (0.15) node[right] {\footnotesize$\rho$};

            \node[below] at (1,-1.5) {$\mathcal{P}_{\text{Hawking}}$};
        \end{scope}
        \draw[rounded corners,fill=black!5] (-3,-1.5) rectangle (5,8);
        \draw[very thick]
            (0,0) coordinate    (M0) .. controls (2,1) and (2,3) ..
            coordinate[pos=0.1] (M1)
            coordinate[pos=0.2] (M2)
            coordinate[pos=0.3] (M3)
            coordinate[pos=0.4] (M4)
            coordinate[pos=0.5] (M5)
            coordinate[pos=0.6] (M6)
            coordinate[pos=0.7] (M7)
            coordinate[pos=0.8] (M8)
            coordinate[pos=0.9] (M9)
            (3,6) coordinate    (M10)
            node[midway,below right] {$C_1$};
        \draw[very thick, shift={(-1,1)}] 
            (0.1,0) coordinate    (N0) .. controls (2.3,0.7) and (1,2.6) ..
            coordinate[pos=0.1] (N1)
            coordinate[pos=0.2] (N2)
            coordinate[pos=0.3] (N3)
            coordinate[pos=0.4] (N4)
            coordinate[pos=0.5] (N5)
            coordinate[pos=0.6] (N6)
            coordinate[pos=0.7] (N7)
            coordinate[pos=0.8] (N8)
            coordinate[pos=0.9] (N9)
            (3,6) coordinate    (N10)
            node[midway,above left] {$C_3$};

        \draw[very thick] (M0) -- (N0) node[midway,below left] {$C_4$};
        \draw[-{latex}] (M1) -- (N1);
        \draw[-{latex}] (M2) -- (N2);
        \draw[-{latex}] (M3) -- (N3);
        \draw[-{latex}] (M4) -- (N4);
        \draw[-{latex}] (M5) -- (N5);
        \draw[-{latex}] (M6) -- (N6);
        \draw[-{latex}] (M7) -- (N7);
        \draw[-{latex}] (M8) -- (N8) node[midway, above] {\footnotesize$V$};
        \draw[-{latex}] (M9) -- (N9);
        \draw[very thick] (M10) -- (N10) node[midway, above right] {$C_2$};

        \fill (0,0) circle (0.15) node[below] {\footnotesize$\phi_0$};
        \fill (3,6) circle (0.15) node[right] {\footnotesize$\phi$};

        \node[below] at (1,-1.5) {$\mathcal{P}_{\text{classical}}$};
        \draw[-{Latex},line width=1.5pt] (4,3.8) .. controls (5,4.2) and (8,4.2) .. (9,3.8) node[midway,above] {\Large$\mathscr{S}$};
    \end{tikzpicture}
    \caption{The curves $C\subset \mathcal{P}_{\text{classical}}$ and $D=\mathscr{S}(C)\subset\mathcal{P}_{\text{Hawking}}$.}
    \label{Figure: F curves}
\end{figure}
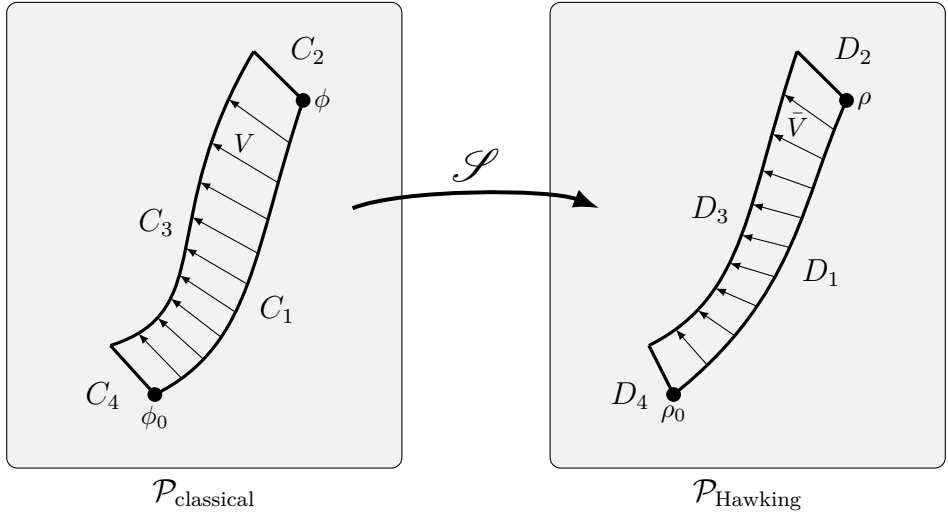

Note that $D$ is the image under $\mathscr{S}$ of a curve of field configurations $C\subset \mathcal{P}_{\text{classical}}$, and using~\eqref{Equation: gamma C Theta} and~\eqref{Equation: Theta I R} we can write
\begin{equation}
    \bar{F}(\rho) = \lim_{\epsilon\to 0}\frac1\epsilon\int_C\Theta
\end{equation}
where $\Theta$ is the symplectic potential of the fields in the island and asymptotic regions. Similarly, each of the four curves $D_1,D_2,D_3,D_4$ are the images under $\mathscr{S}$ of four curves $C_1,C_2,C_3,C_4$ respectively, where $C_1$ connects $\phi_0$ to $\phi$ (which obey $\mathscr{S}(\phi)=\rho$, $\mathscr{S}(\phi_0)=\rho_0$), and we have
\begin{equation}
    C_2 = \{\psi_t(\phi):t\in [0,\epsilon]\}, \quad C_3 = \psi_\epsilon(C_1), \quad C_4 = \{\psi_t(\phi_0):t\in [0,\epsilon]\},
\end{equation}
where $\psi_t$ is the flow generated by the vector field $V$. Both of the curves $C$ and $D$ are depicted in Figure~\ref{Figure: F curves}.

Taking into account the correct orientations of the components of $C$, we have
\begin{nalign}
    \bar{F}(\rho) &= \lim_{\epsilon\to 0}\frac1\epsilon\qty[\int_{C_1}\Theta + \int_{C_2}\Theta + \int_{C_3}\Theta + \int_{C_4}\Theta]\\
    &= \partial_\epsilon\qty[\int_0^\epsilon\dd{s}(\iota_V\Theta)(\psi_s(\rho)) - \int_{\psi_\epsilon(C_1)} \Theta - \int_0^\epsilon\dd{s}(\iota_V\Theta)(\psi_s(\rho_0))]_{\epsilon=0} \\
    &= (\iota_V\Theta)(\rho) - \underbrace{\int_{C_1}\lie_V\Theta}_{\mathclap{=\delta(\iota_V\Theta) + \iota_V\delta\Theta}} - (\iota_V\Theta)(\rho_0) \\
    &= -\int_{C_1}\iota_V\Omega_{I\cup R},
\end{nalign}
where $\Omega_{I\cup R}=\delta\Theta$. But now, using~\eqref{Equation: tilde F hamiltonian} and~\eqref{Equation: V vector I R}, we may write 
\begin{equation}
    \bar{F}(\rho) = \int_{D_1} \delta{\tilde{F}} = \tilde{F}(\phi) - \tilde{F}(\phi_0).
\end{equation}
Thus 
\begin{equation}
    \bar{F}(\rho_i) - \bar{F}(\rho_j) = \tilde{F}(\phi_i)-\tilde{F}(\phi_j) = F(\phi_i) - F(\phi_j).
\end{equation}

Therefore, by the lemma in the previous section, the information in the diary is recoverable from the Hawking radiation. Moreover, we have described the protocol by which one recovers the information: one picks a function $F$ of the classical information, one finds the Hamiltonian vector $V$ that it generates, one pushes this vector forward to the space of Hawking states, one uses it to construct a curve of states, and then one computes the Uhlmann phase around that curve. The result is a reconstruction of the value of the function $F$, in the way just described.

\subsection{Island phase space reconstruction}

Our result also makes it possible to recover the full structure of the island phase space -- so in particular we can compute Poisson brackets of all the functions of the information in the black hole. By measuring the Uhlmann phases of various curves in $\mathcal{P}_{\text{Hawking}}$, one can completely recover all of the components of the symplectic form $\Omega_{I\cup R}$. Actually, because all the configurations in $\mathcal{P}_{\text{classical}}$ are assumed to lead to the same classical exterior, in this case $\Omega_{I\cup R}$ is just the symplectic form for the fields on the island $I$ (because $\phi$ will be constant in $R$, so $\delta\phi|_R=0$).

In this way, although $\mathcal{P}_{\text{Hawking}}$ is a space of quantum states, in the semiclassical limit it aquires a symplectic structure. Thus, it becomes a symplectic manifold, and so (in the semiclassical limit) can be thought of as a classical phase space. As a phase space it is isomorphic to the phase space of the fields on the island. So we have reconstructed the classical phase space of the island: it is just isomorphic to $\mathcal{P}_{\text{Hawking}}$ with the symplectic form $\mathscr{S}^*\Omega_{I\cup R}$. This means we are able to compute arbitrary Poisson brackets of fields on the island by measuring the Hawking radiation Uhlmann phase.

Note that we do not need to know anything about the map $\mathscr{S}$ in order to construct Poisson brackets of island observables. In particular, suppose $\bar{F}_1,\bar{F}_2$ are two functions on $\mathcal{P}_{\text{Hawking}}$. These can be viewed as classical observables, and indeed this is clear if we pull them back through $\mathscr{S}$:
\begin{equation}
    F_1 = \bar{F}_1\circ\mathscr{S}, \quad F_2 = \bar{F}_2 \circ\mathscr{S}.
    \label{Equation: hawking classical observable}
\end{equation}
If we don't know about $\mathscr{S}$, then we can't directly access $F_1,F_2$. However, using the Uhlmann phase, we can deduce all the components of the symplectic form on $\mathcal{P}_{\text{Hawking}}$, and we can do this \emph{without} needing to use $\mathscr{S}$. We can use this symplectic form to compute the Poisson bracket of $\bar{F}_1$ and $\bar{F}_2$
\begin{equation}
    \pb{\bar{F}_1}{\bar{F}_2}_{\text{Hawking}} = -(\mathscr{S}^*\Omega_{I\cup R})^{-1}(\delta\bar{F}_1,\delta\bar{F}_2).
\end{equation}
As in~\eqref{Equation: hawking classical observable}, this corresponds to the classical observable
\begin{nalign}
    \pb{\bar{F}_1}{\bar{F}_2}_{\text{Hawking}}\circ \mathscr{S} &= -(\mathscr{S}^*\Omega_{I\cup R})^{-1}(\delta\bar{F}_1,\delta\bar{F}_2)\circ S \\
                                                                &= -\Omega_{I\cup R}^{-1}(\mathscr{S}^*\delta \bar{F}_1,\mathscr{S}^*\delta \bar{F}_2)\\
                                                                &= -\Omega_{I\cup R}^{-1}(\delta F_1,\delta F_2) \\
                                                                &= \pb{F_1}{F_2}_{\text{classical}}.
        \label{Equation: Poisson no S}
\end{nalign}
Thus, we can construct classical Poisson brackets using the Uhlmann phase alone, without knowledge of $\mathscr{S}$. We only need to know about $\mathscr{S}$ when we want to actually evaluate these Poisson brackets on a given Hawking radiation state. 

Finally, if we wait long enough after the Page time, the island will grow to span the entire black hole interior. At that point, we can use our results to recover all information that fell inside the black hole, and to reconstruct the entire classical phase space of the interior.

\section{Conclusion}
\label{Section: Conclusion}

\begin{figure}
    \centering
    \begin{tikzpicture}[thick,scale=0.9]
        \draw[fill=black!5] (0,0) ellipse (3 and 2);
        \node at (0,0.5) {\Large$\mathcal{P}_{\text{classical}}$};
        \node at (0,-0.5) {symplectic form $\Omega$};
        \draw[fill=black!5] (9,0) ellipse (3 and 2);
        \node at (9,0.8) {\Large$\mathcal{P}_{\text{Hawking}}$};
        \node at (9,-0.2) {symplectic form $\Omega_{I\cup R}$};
        \node at (9,-0.8) {(from Uhlmann phase)};

        \draw[-{Latex},line width=1.5pt] (2.5,1.8) .. controls (3.2,2.2) and (5.8,2.2) .. (6.5,1.8) node[midway,above] {\Large$\mathscr{S}$};
        \node at (4.5,3.3) {semiclassical evolution};
        \draw[{Latex}-,line width=1.5pt,dashed] (2.5,-1.8) .. controls (3.2,-2.2) and (5.8,-2.2) .. (6.5,-1.8) node[midway,below] {\Large$\mathscr{S}^{-1}$};
        \node at (4.5,-3.3) {(after a sufficiently long time)};
    \end{tikzpicture}
    \caption{We reconstruct the island phase space by measuring the Uhlmann phase of the Hawking radiation. After a sufficiently long time, the map $\mathscr{S}$ representing the evolution between the space of classical fields and the space of Hawking radiation states becomes invertible, and all classical information is recovered.}
    \label{Figure: big picture}
\end{figure}
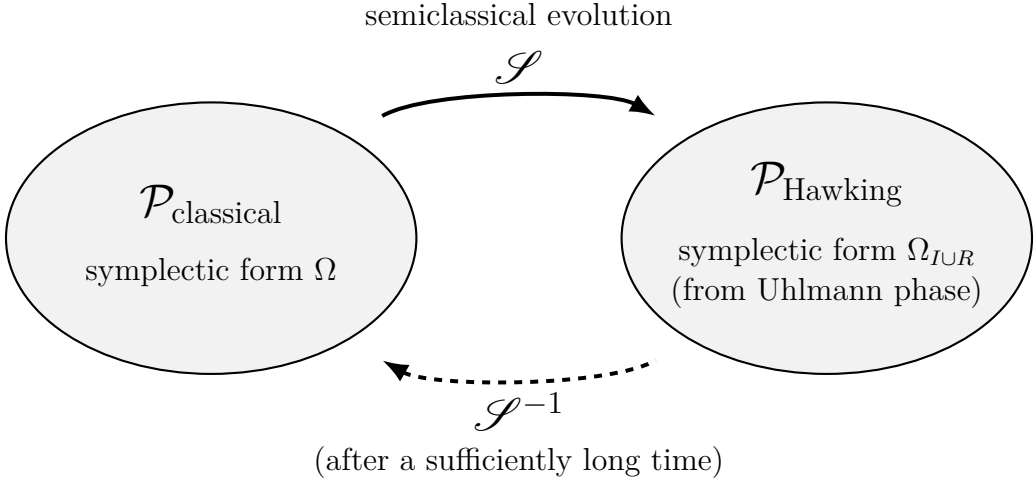

In this paper, we have shown that the curvature of the Uhlmann phase of Hawking radiation is equivalent to the symplectic form of the island and the radiation. This calculation depended in a crucial way on the contributions of replica wormholes to the gravitational path integral. Our results have provided a protocol for reconstructing the phase space of the black hole interior. All classical information is explicitly recovered after enough Hawking radiation is emitted. This is depicted in Figure~\ref{Figure: big picture}.

We have so far restrained from discussing the operational viability of our information recovery protocol. In particular, we have assumed that as an observer we have access to some experimental apparatus that allows us to move the state of the Hawking radiation around certain curves in $\mathcal{P}_{\text{Hawking}}$, and to measure the resulting Uhlmann phase. It is clear that the apparatus only needs access to the state of the Hawking radiation, so in principle interior information can be recovered from the outside of the black hole, but we have not considered what kind of computational resources we would need in order to construct and use the apparatus. We will leave this important question to future work, but it is worth making two comments. 

First, any process which recovers information from Hawking radiation is believed to be highly algorithmically complex~\cite{Complex}, and this should apply to the results of this paper. There are two stages to our protocol: the construction of the curve of radiation states, and the measurement of the Uhlmann phase around that curve. It seems highly likely that for our protocol the complexity comes in at the first stage, because the construction of the curve uses the dynamical map $\mathscr{S}:\mathcal{P}_{\text{classical}}\to\mathcal{P}_{\text{Hawking}}$. The most obvious way to experimentally implement this map would be to simulate the evolution of the black hole on a quantum computer, which is a very complex procedure. On the other hand, there is no obvious reason why the measurement of the Uhlmann phase itself should be complex, once we know which curve of states to measure it around. As shown in~\eqref{Equation: Poisson no S}, we do not need $\mathscr{S}$ to access the Poisson brackets of classical observables. So if we are only interested in the algebraic relations of observables in the black hole interior, we don't need to simulate the evolution of the black hole, i.e.\ we can avoid the complex stage of the protocol.

Second, there is an important distinction between `asymptotic' and `one-shot' information recovery. In the former case, we are tasked with recovering information using an arbitary number of copies of a final state. In the latter case, we only have a single copy of the final state from which we need to be able to recover the information with high probability. One-shot information recovery is strictly more difficult than asymptotic information recovery, but it is also strictly more desirable in a practical setting. There exists in the literature a description of an experiment for measuring Uhlmann phase~\cite{Operational1,Operational2}, but this measurement is an asymptotic one, because it requires many copies of the state. The distinction between the asymptotic and one-shot types of information recovery is particularly relevant to quantum systems, because the no-cloning theorem implies that there is no way of producing an arbitary number of copies of the final state, if we start with just a single copy. However, even though we must do quantum measurements to find the Uhlmann phase, our end goal in the present work is only the recovery of classical information. There is no problem with copying classical information an arbitrary number of times with arbitrary accuracy in the classical limit. Our expectation therefore is that if there is an asymptotic implementation of the protocol described here, then there is also a one-shot implementation. It would be very interesting to see whether this expectation is correct.

We should mention that another method for recovering black hole interior information from the Hawking radiation uses the so-called `Petz map', as described in~\cite{Flurry2}. This map provides a way to reconstruct operators acting on the fields in the island as operators acting on the Hawking radiation. In this way it provides a way to recover \emph{quantum} information (in the form of expectation values of operators acting inside the black hole), which is in contrast to the \emph{classical} information recovery described in this work. It is clear that the protocol described here is quite different to the one involving the Petz map. Note that not all classical observables can be represented as the expectation values of quantum operators -- so it is plausible that our protocol provides access to some information that the Petz map does not. On the other hand, in the classical limit, the two protocols may become more closely aligned, and this would be interesting to check. It would also be interesting to understand how the Petz map and Uhlmann phase are related in a more general setting.

Before ending the paper, we want to just point out an interesting consequence of our results. We have been fairly vague about the type of theory we are considering in the asymptotic region. Really, the only condition it has to satisfy is that it can be consistently coupled to the gravitational theory at the cutoff surface, in such a way that it absorbs the Hawking radiation.

But what we want to claim is that \emph{any} theory which can be consistently coupled to gravity in this way \emph{is itself a gravitational theory}, at least in some kind of classical limit.

To see this, let us take a state in such a theory, and couple it to gravity. We then form a black hole in the gravitational region, and let it evaporate. Once the asymptotic theory has absorbed the Hawking radiation, we throw away the gravitational theory -- we don't care about it any more. We are only interested in the state of the asymptotic theory $\rho_R$.

As we have shown, the curvature of the Uhlmann phase of such states is a symplectic form for gravitational fields. So even though we have discarded the gravitational theory, we still have a class of states in the asymptotic theory with a gravitational interpretation. If we are interested in a classical limit in which the Uhlmann phase is important, then this necessarily \emph{implies} that the classical limit of the asymptotic theory is a theory of gravity.

We think this observation has two possible explanations. Either: the only theories in the asymptotic region which can be consistently coupled to the gravitational region must already have had gravity in them. Or: there is a `gravitational sector' of density matrices in a wide class of theories that don't a priori contain gravity in a traditional sense. Both of these possibilities seem interesting, and we leave their exploration to future work.

\section*{Acknowledgements}
\addcontentsline{toc}{section}{\protect\numberline{}Acknowledgements}

I wish to thank Philipp Hoehn, Isha Kotecha and Fabio Mele for helpful discussions and comments. This work was supported in part by the Okinawa Institute of Science and Technology. I am also grateful for the continued hospitality of the University of Cambridge while I am stuck in the UK.

\printbibliography

\end{document}